\documentclass[preprint,aps,pra,showpacs,floatfix]{revtex4}
\usepackage[dvips]{graphicx}

\usepackage{longtable}
\usepackage{dcolumn}
\usepackage[dvips]{graphicx}
\usepackage{bm}
\usepackage{bbm}

\usepackage{times}
\usepackage{nicefrac}
\usepackage{amsmath}
\usepackage{amsfonts}
\usepackage{amssymb}
\usepackage{amsthm}

\newcolumntype{.}{D{x}{}{-1}}

\begin{document}
%%%%%%%%%%%%%%%%%%%%%%%%%%%%%%%%%%%%%%%%%%%%%%%%%%%%%%%%%%

\newcommand{\vare}{\varepsilon}

\newcommand{\pr}{^{\prime}}
\newcommand{\bfx}{{\bf x}}
\newcommand{\bfy}{{\bf y}}
\newcommand{\bfz}{{\bf z}}
\newcommand{\bfp}{{\bf p}}
\newcommand{\la}{\langle}
\newcommand{\ra}{\rangle}
\newcommand{\eps}{\varepsilon}
\newcommand{\beq}{\begin{equation}}
\newcommand{\eeq }{\end{equation}}
\newcommand{\beqn}{\begin{eqnarray}}
\newcommand{\eeqn }{\end{eqnarray}}
\newcommand{\ba}{\begin{array}}
\newcommand{\ea}{\end{array}}
\newcommand{\balpha}{{\mbox{\boldmath$\alpha$}}}
\newcommand{\az}{\alpha Z}
\newcommand{\aZ}{\alpha Z}
\newcommand{\etal}{{\it et al. }}

\newcommand{\lbr}{\langle}
\newcommand{\rbr}{\rangle}

\newcommand{\Dmatrix}[4]{
        \left(
        \begin{array}{cc}
        #1  & #2   \\
        #3  & #4   \\
        \end{array}
        \right)
        }

%%%%%%%%%%%%%%%%%%%%%%%%%%%%%%%%%%%%%%%%%%%%%%%%%%%%%%%%%%
\title{QED calculation of the n=1 and n=2 energy levels in He-like ions}
\author{A.~N.~Artemyev,$^{1,2}$ V.~M.~Shabaev,$^{1,2}$
  V.~A.~Yerokhin,$^{1,2,3}$ G. Plunien,$^{2}$  and G.Soff$^{2}$}
\affiliation{$^{1}$Department of Physics, St. Petersburg State University,
Oulianovskaya 1, Petrodvorets, St. Petersburg 198504, Russia\\
$^2$Institut f\"{u}r Theoretische Physik, TU Dresden, Mommsenstra{\ss}e 13,
D-01062 Dresden, Germany\\
$^3$ Center for Advanced Studies, St. Petersburg State Polytechnical
University, Polytekhnicheskaya 29, St. Petersburg 195251, Russia
}

\begin{abstract}

We perform {\it ab initio} QED calculations of energy levels for the $n=1$
and $n=2$ states of He-like ions with the nuclear charge in the range $Z =
12$-$100$. The complete set of two-electron QED corrections is evaluated to
all orders in the parameter $\aZ$. Uncalculated contributions to energy
levels come through orders $\alpha^3 (\aZ)^2$, $\alpha^2 (\aZ)^7$, and
higher. The calculation presented is the first treatment for excited states
of He-like ions complete through order $\alpha^2 (\aZ)^4$. A significant
improvement in accuracy of theoretical predictions is achieved, especially in
the high-$Z$ region.

\end{abstract}

\maketitle

%%%%%%%%%%%%%%%%%%%%%%%%%%%%%%%%%%%%%%%%%%%%
\section*{Introduction}

Helium and helium-like ions, being the simplest many-electron systems,
traditionally serve as an important testing ground for investigations of
many-body relativistic and QED effects. Calculations of QED effects in
He-like ions have a long history. The expression for the Lamb shift complete
through orders $\alpha (\aZ)^4$ and $\alpha^2 (\aZ)^3$ was derived in
pioneering studies by Araki \cite{araki:57} and Sucher \cite{sucher:58}.
Numerous posterior investigations of higher-order QED corrections in
two-electron systems (see, e.g., review \cite{drake:01:hydr} and recent
original studies \cite{pachucki:00:prl,korobov:01,pachucki:02:jpb}) were
primarily aimed at helium, in which the experimental accuracy is by far
better than in other two-electron systems. Recent progress in experimental
spectroscopy of highly charged ions
\cite{marrs:95,stoehlker:96,gumberidze:04} opened new perspectives for
probing higher-order QED effects in ions along the helium isoelectronic
sequence up to He-like uranium. Investigations of QED effects in high-$Z$
ions are of particular importance since they can provide tests of quantum
electrodynamics in the region of a very strong Coulomb field of the nucleus.
Another factor that stimulates these investigations is the possibility to
test the standard model by studying the effects of parity non-conservation
(PNC) \cite{schaefer:89,karasiev:92,maul:96,labzowsky:01}. Experimental
identification of the PNC effects will require precise knowledge of the
$2^1S_0-2^3P_0$ interval in He-like ions with nuclear charge numbers near
$Z=64$ (gadolinium) and $Z=90$ (thorium), which happens to be very small for
these values of $Z$ thus enhancing the PNC effects significantly.

Investigations of QED effects in heavy He-like ions differ significantly
from those for the helium atom. First of all, the nuclear coupling parameter
$\aZ$ approaches unity and cannot be regarded as a good expansion
parameter as in the case of helium. But on the
other side, the electron-electron interaction in these systems is suppressed
by a factor of $1/Z$ with respect to the electron-nucleus interaction and,
therefore, can be accounted for by a perturbation expansion in the
parameter $1/Z$.

Until recently, the only QED effects calculated to all orders in $\aZ$ were
the one-electron self-energy and vacuum-polarization corrections
\cite{mohr:74,soff:88:vp}. So, theoretical investigations of energy levels in
heavy He-like ions mostly relied on these one-electron values, correcting
them to account for the ``screening'' effect by various semi-empirical rules,
notably, within Welton's approximation, as in Ref.~\cite{indelicato:87}. A
more elaborate treatment of QED effects in He-like ions was presented by
Drake \cite{drake:CJP:88}. His values for the QED correction included the
complete contribution to order $\alpha^2(\aZ)^3$ derived in
Refs.~\cite{araki:57,sucher:58} and parts of higher-order contributions
obtained by employing the all-order results available for the one-electron
QED corrections. The total energy values of Ref.~\cite{drake:CJP:88} are
complete through order $\alpha^2(\aZ)^3$ and uncalculated terms start in
orders $\alpha^2(\aZ)^4$ and $\alpha^3(\aZ)^2$.

Later, Johnson and Sapirstein \cite{johnson:92} applied relativistic
many-body perturbation theory (MBPT) to the treatment of the electron
correlation for $n=2$ triplet states of He-like ions. Combined with Drake's
values for the QED and recoil corrections, their results yielded a better
agreement with the experimental data than those of Ref.~\cite{drake:CJP:88}.
While the approach of Ref.~\cite{johnson:92} is still incomplete to order
$\alpha^2(\aZ)^4$, it includes terms that were not accounted for in
Ref.~\cite{drake:CJP:88}, namely the Breit-Breit interaction and some
relativistic corrections to the second-order energy. Later, other evaluations
of the electron-correlation part of the energies of He-like ions were
performed by the relativistic configuration-interaction (CI) method
\cite{chen:93:pra} and by the relativistic all-order MBPT approach
\cite{plante:94}. The studies \cite{johnson:92,chen:93:pra,plante:94} share
the same main features: their treatment is based on the no-pair Hamiltonian
and the electron correlation is taken into account within the Breit
approximation. The results of these evaluations are in a very good agreement
with each other.

A somewhat different approach was employed in
Refs.~\cite{cheng:94:pra,cheng:00}. While the electron-correlation part was
evaluated (as in the previous work by the same group \cite{chen:93:pra}) by
the CI method, the QED part was not taken from Ref.~\cite{drake:CJP:88} but
evaluated independently, by considering the one-loop QED corrections in a
local screening potential. Due to different treatments of QED effects, there
are certain deviations between the results of
Refs.~\cite{cheng:94:pra,cheng:00} and those of
Refs.~\cite{johnson:92,chen:93:pra,plante:94}.

In order to obtain reliable predictions for energy levels of high-$Z$ ions
and to improve the theoretical accuracy in the low- and middle-$Z$ region, it
is necessary to take into account two-electron QED effects without an
expansion in $\aZ$. Such project has been recently accomplished (up to order
$\alpha^2$) for the two-electron part of the ground-state energy of He-like
ions
\cite{blundell:93:b,lindgren:95:pra,persson:96:2el,artemyev:97,yerokhin:97:pla}
and for the lowest-lying states of Li-like ions
\cite{artemyev:99,yerokhin:99:sescr,yerokhin:00,sapirstein:01:lamb}. 
To perform similar QED
calculations for excited states of He-like ions is more difficult. One of the
reasons is that, for the first time in QED calculations to all orders in
$\aZ$, we encounter levels that are quasi-degenerate, namely $2^3P_1$ and
$2^1P_1$. To derive formal expressions for QED corrections in case of
quasi-degenerate states is a serious problem that has been solved first
within the two-time Green function (TTGF) method
\cite{shabaev:90:tmf,shabaev:93,shabaev:02:rep}. Different approaches to this
problem have recently been addressed by other authors
\cite{lindgren:01,andreev:04}.

Several QED corrections have been calculated to all orders in $\aZ$ for
excited states of He-like ions up to now. In our previous investigation
\cite{artemyev:00:pra}, we evaluated the vacuum-polarization screening
correction for all $n=2$ states of He-like ions. The two-photon exchange
correction was calculated for excited states of He-like ions by Mohr and
Sapirstein \cite{mohr:00:pra} ($2^3S_1$ and $2^3P_{0,2}$ states), by Andreev
\etal \cite{andreev:01,andreev:03} ($2^1S_{0,1}$, $2^3P_0$) and
\cite{andreev:04} ($2^{1,3}P_1$), and by \r{A}sen \etal \cite{asen:02}
($2^1S_{0,1}$). In this paper we present an evaluation of the self-energy
screening correction and an independent calculation of the two-photon
exchange correction for all $n=2$ states of He-like ions. This completes the
{\it ab initio} treatment of all {\it two-electron} QED corrections of order
$\alpha^2$ to all orders in $\aZ$ and significantly improves the theoretical
accuracy for the energy values, especially in the high-$Z$ region. Unlike all
previous calculations, the results obtained are complete through order
$\alpha^2(\aZ)^4$; uncalculated terms enter through three-photon QED effects
(to order $\alpha^3(\aZ)^2$ and higher) and through two-loop
one-electron QED corrections ($\alpha^2(\aZ)^7$ and higher).

The paper is organized as follows. In the next section we describe the basic
formalism and present general formulas for the two-electron QED corrections
for the case of quasi-degenerate levels. In Section~\ref{sec:twoelQED},
the numerical procedure is briefly discussed and numerical results
are presented for the two-photon exchange correction and the screened self-energy
correction. The total two-electron QED correction is then compiled, analyzed,
and compared with the known terms of the $\aZ$ expansion. In the last section,
we present a compilation of all contributions available to the energy levels and
compare results of different theoretical evaluations with existing
experimental data. The relativistic units ($\hbar=c=m=1$) are used throughout
the paper.

%%%%%%%%%%%%%%%%%%%%%%%%%%%%%%%%%%%%%%%%%%%%%%%%%%%%%%%%%%%%%%%%%%%%%%%%%%%%
%
%
%%%%%%%%%%%%%%%%%%%%%%%%%%%%%%%%%%%%%%%%%%%%%%%%%%%%%%%%%%%%%%%%%%%%%%%%%%%%
\section{Formal expressions}
\label{sec:derivation}

%%%%%%%%%%%%%%%%%%%%%%%%%%%%%%%%%%%%%%%%%%%%%%%%%%%%%%%%%%%%%%%%%%%%%%%%%%%%
\subsection{Basic formalism}

In this section we briefly formulate the basic equations of the TTGF method
for quasi-degenerate  states of a He-like ion. A detailed description of the
method and, particularly, its implementation for the case of quasidegenerate
states can be found in
Refs.~\cite{shabaev:93,shabaev:94:ttg2,shabaev:02:rep}. The derivation will
be given for two particular quasidegenerate states, $(1s2p_{1/2})_1$ and
$(1s2p_{3/2})_1$, and can immediately be extended to a more general case. The
unperturbed two-electron wave functions in the $jj$ coupling are given by
\beqn
\label{u1}
 u_1&=&\sum_{m_a m_v}\langle j_a m_a j_v m_v|JM\rangle
    \frac{1}{\sqrt{2}}\sum_{P}(-1)^P |PaPv\rangle\,,\\
 \label{u2}
 u_2&=&\sum_{m_a
    m_w}\langle j_a m_a j_w m_w|JM\rangle \frac{1}{\sqrt{2}}\sum_{P}(-1)^P
|PaPw\rangle\,,
\eeqn
where $a$, $v$ and $w$ are taken to represent $1s$, $2p_{1/2}$ and $2p_{3/2}$
orbitals, respectively; $P$ is the permutation operator: $$ \sum_{P}(-1)^P
|PaPv\rangle=|av\rangle-|va\rangle\,, $$ and $|av\rangle \equiv |a\rangle
|v\rangle$ is the product of the one-electron Dirac wave functions. The
transition to the wave functions corresponding to the LS-coupling scheme
within the non-relativistic approximation can be performed by
\beq
\left( {|2^3P_1\rbr}\atop{|2^1P_1 \rbr} \right) = R \left(
{|(1s2p_{1/2})_1\rbr}\atop{|(1s2p_{3/2})_1 \rbr} \right)\,,
\eeq
with
\beq    \label{Rmatrix}
R  = \frac1{\sqrt{3}}\Dmatrix{\sqrt{2}}{-1}{1}{\sqrt{2}}\,.
\eeq
We mention that this choice of the matrix $R$ implies that the
one-electron $2p_{1/2}$ and $2p_{3/2}$ wave functions have the same sign in
the non-relativistic limit.

The standard definition of the four-time two-electron Green function in the
external field of the nucleus is
\beqn  \label{eq1} G(x_1',x_2';x_1,x_2)=\la
0|T\psi(x_1')\psi(x_2')\bar\psi(x_1) \bar\psi(x_2)|0\ra\,,
\eeqn
where $\psi(x)$ is the electron-positron field operator in the Heisenberg
representation, $\bar\psi=\psi^{\dag}\gamma^0$, and $T$ denotes the
time-ordered product operator. This Green function is constructed by
perturbation theory after the transition to the interaction representation
where it is given by (see, e.g., \cite {itzykson:80})
\beqn  \label{Gint}
G(x_{1}^{\prime},x_{2}^{\prime};x_{1},x_{2})
%   \nonumber\\ & &
   =
\frac{ \langle 0|T\psi_{\rm in}(x_{1}^{\prime})\psi_{\rm
in}(x_{2}^{\prime})\overline{\psi }_{\rm in} (x_{2})\overline{\psi}_{\rm
in}(x_{1})\exp\left[-i\int d^{4}z{\cal H}_{\rm int} (z)\right]
|0\rangle}{\langle 0|T\exp\left[-i\int d^{4}z{\cal H}_{\rm int} (z)\right]
|0\rangle}\,.
\eeqn
Here $\psi_{\rm in}(x)$ is the electron-positron field operator in the
interaction representation and ${\cal H}_{\rm int}$ is the interaction
Hamiltonian. Expression (\ref{Gint}) allows one to construct $G$ by
using Wick's theorem.

In what follows, it is more convenient to work with the Green function in the
mixed energy-coordinate representation, which is defined by
\beqn
\label{gmix} G(p'^0_1,\bfx_{1}', p'^0_2, \bfx_{2}';p^0_1,\bfx_1,
p^0_2,\bfx_2) &=&\frac{1}{(2\pi)^4}\int^{\infty}_{-\infty}
        dx_1^0\,dx_2^0\,dx'^0_1\, dx'^0_2\, \nonumber \\
&&\times\exp{(ip'^0_1x'^0_1+ip'^0_2x'^0_2-ip_1^0x_1^0-ip_2^0x_2^0)}
\;G(x_1',x_2';x_1,x_2)\,.
    \nonumber\\
\eeqn
The Feynman rules for $G(p'^0_1,\bfx_{1}', p'^0_2, \bfx_{2}';p^0_1,\bfx_1,
p^0_2,\bfx_2)$ can be found in \cite{shabaev:94:ttg2,shabaev:02:rep}.
We now introduce the Green function $g(E)$ as
\beqn
    \label{g}
 g(E)\,\delta(E-E')&=&\frac \pi i \int^{\infty}_{-\infty}
dp_1^0\,dp_2^0\,dp'^0_1\, dp'^0_2\, \delta(E-p_1^0-p_2^0)
 \nonumber \\ &&\times
 \delta(E'-p'^0_1-p'^0_2)\,
P_0\,G(p'^0_1,p'^0_2;p^0_1,p^0_2)\,\gamma_1^0\gamma_2^0\,P_0\,,
\eeqn
where $P_0=\sum\limits_k u_k u^{\dag}_k$ is the projector on the subspace of
the unperturbed quasi-degenerate states under consideration [see
Eqs.~(\ref{u1}) and (\ref{u2})]. It can easily be shown (see, e.g.,
Refs.~\cite{shabaev:94:ttg2,shabaev:02:rep}) that the Green function $g$ is
the Fourier transform of the two-time Green function projected on the
subspace of the unperturbed quasi-degenerate states.

It can be derived (see Ref.~\cite{shabaev:02:rep} for details) that the
system under consideration can be described by a two-dimensional
Schr\"odinger-like equation ($k = 1,2$),
\beqn  \label{schr}
 H\psi_k=E_k\psi_k\,,
\ \ \ \ \
 \psi_k^{\dag}\,\psi_{k'}=\delta_{kk'}\,,
\eeqn
where
\beqn  \label{schr0}
  H&=&P^{-1/2}KP^{-1/2}\,,\\
  K&=&\frac{1}{2\pi i}\oint_\Gamma dE\, E\, g(E)\,,\\
  P&=&\frac{1}{2\pi i}\oint_\Gamma dE\, g(E)\,,
\eeqn
$\Gamma$ is a contour in the complex $E$ plane that surrounds the levels
under consideration but does not encircles other levels, and $E_k$  are the
exact energies of the states under consideration. It is assumed that the
contour $\Gamma$ is oriented anticlockwise. The operator $H$, which is
a $2\times2$ matrix, is constructed by perturbation theory in $\alpha$.
Substituting
\beqn
g(E)&=&g^{(0)}(E)+g^{(1)}(E)+g^{(2)}(E)+\cdots\,,\\
P&=&P^{(0)}+P^{(1)}+P^{(2)}+\cdots\,,\\ K&=&K^{(0)}+K^{(1)}+K^{(2)}+\cdots\,,
\eeqn
where the superscript indicates the order in $\alpha$, we obtain \cite{shabaev:93}
\beq
H^{(0)}=K^{(0)}\,,
\eeq
\beq
 \label{H1} H^{(1)}=K^{(1)}-\frac 12
P^{(1)}K^{(0)}-\frac 12 K^{(0)}P^{(1)}\,,
\eeq
\beqn \label{H2}
H^{(2)}&=&K^{(2)}-\frac 12 P^{(2)}K^{(0)}-\frac 12 K^{(0)}P^{(2)}-\frac 12
P^{(1)}K^{(1)}-\frac 12 K^{(1)}P^{(1)}
 \nonumber \\ &&
 +\frac 38 P^{(1)}P^{(1)}K^{(0)}+\frac 38 K^{(0)}P^{(1)}P^{(1)}+\frac 14
P^{(1)}K^{(0)}P^{(1)}\,.
\eeqn
The solvability of Eq.~(\ref{schr}) yields the basic equation for the
calculation of the energy levels
\beq
\label{schr1} {\rm det}(E-H)=0\,.
\eeq
As was noticed in Ref.~\cite{shabaev:93}, due to nonzero decay rates of
excited states, the self-adjoint part of $H$ should be understood in
Eqs.~(\ref{schr}) and (\ref{schr1}),
\beq
 H\equiv (1/2)(H+H^{\dag})\,.
\eeq

To zeroth order in $\alpha$, the Green function $g(E)$ is
\beqn
\label{g0} g^{(0)}(E)=\sum_{s=1}^{2}\frac{|u_{s}\rangle \langle
u_{s}|}{E-E_{s}^{(0)}}\,,
\eeqn
where $E_{1}^{(0)}$ and  $E_{2}^{(0)}$ are the unperturbed energies of the
$(1s2p_{1/2})_1$ and $(1s2p_{3/2})_1$ states, respectively, given by the sum
of the one-electron Dirac-Coulomb energies: $$
E_1^{(0)}=\varepsilon_{1s}+\varepsilon_{2p_{1/2}}\,,\;\;\;\;\;\;\;\;\;\;
E_2^{(0)}=\varepsilon_{1s}+\varepsilon_{2p_{3/2}}\,. $$ Substituting
Eq.~(\ref{g0}) into the definitions of $K$, $P$, and $H$, one gets
\beqn
K^{(0)}_{ik}&=&E^{(0)}_{i}\delta_{ik}\,,\\ P^{(0)}_{ik}&=&\delta_{ik}\,,\\
H^{(0)}_{ik}&=&E^{(0)}_{i}\delta_{ik}\,.
\eeqn

Now we introduce a set of notations that will shorten the following
expressions. The short-hand notation will be used for the summation over the
Clebsh-Gordan coefficients in Eqs.~(\ref{u1}), (\ref{u2}):
\beq \label{Fi}
F_i\, |i_1 i_2\rangle \equiv\sum_{m_{i_1} m_{i_2}}\langle j_{i_1}m_{i_1}
j_{i_2} m_{i_2}|JM \rangle\, |i_1 i_2\rangle\,.
\eeq
where $|i_1i_2\rangle$ is either $|av\rangle$ or $|aw\rangle$. It is
convenient also to use the notation for the operator of the electron-electron
interaction:
 \begin{equation}
      I(\omega) =  e^2\, \alpha_{1}^{\mu}\, \alpha_{2}^{\nu}\, D_{\mu\nu}
      (\omega)\,,
    \label{I}
\end{equation}
where $\alpha^{\mu} = \gamma^{0}\gamma^{\mu}= (1,\mbox{\boldmath $\alpha$})$
and $D_{\mu\nu}$ denotes the photon propagator. In the Feynman gauge, the
propagator of a photon with the non-zero mass $\mu$ is
  \begin{equation}
 D_{\mu\nu} (\omega,{\bf x}-{\bf y})=
         g_{\mu\nu}\frac{\exp{(i\sqrt{\omega^{2}-\mu^2+i0}\;|
        {\bf x}-{\bf y}|)}}
         {4\pi|{\bf x}-{\bf y}|}\,,
                 \label{D}
\end{equation}
where it is assumed that $ {\rm Im}{\sqrt{\omega^{2}-\mu^2+i0}}>0 $.
For the matrix elements of the
operator $I(\omega)$ we will use the short-hand notation
\beqn
I_{ijkl}(\omega)= \langle ij|I(\omega)|kl\rangle\,.
\eeqn

%%%%%%%%%%%%%%%%%%%%%%%%%%%%%%%%%%%%%%%%%%%%%%%%%%%%%%%%%%%%%%%%%%%%%%%%%
\subsection{One-photon exchange diagram}

In order to illustrate how the method works, below we present the detailed
derivation of the correction to the quasidegenerate energy levels
$(1s2p_{1/2})_1$ and $(1s2p_{3/2})_1$ due to the one-photon exchange diagram
(Fig.~\ref{fig:onephot}). While the corresponding evaluation is much less
cumbersome than those for the second-order two-electron corrections, it
demonstrates most essential features that are encountered in these cases. For
simplicity, in the derivation below we will assume that the unperturbed
energy of the initial state $i$ differs from that of the final state $k$:
$E_{i}^{(0)}\not =E_k^{(0)}$ (in the case under consideration it corresponds
to $i\not = k$). However, all the final formulas can be shown to be valid
also for the case $E_{i}^{(0)}=E_k^{(0)}$.

According to the Feynman rules \cite{shabaev:94:ttg2,shabaev:02:rep} and the
definition of $g(E)$, the contribution of the one-photon exchange diagram is
\beqn
             g_{ik}^{(1)}(E)&=&F_i F_k
     \Bigl(\frac{i}{2\pi}\Bigr)^{2}
        \int_{-\infty}^{\infty}dp_{1}^{0}\,dp_{1}^{\prime 0}\,
                           \sum_{P}(-1)^{P}
             \frac{1}{(p_{1}^{\prime 0}-\varepsilon_{P i_1}+i0)
                (E-p_{1}^{\prime 0}-\varepsilon_{P i_2}+i0)}
 \nonumber \\ && \times
             \frac{I_{P i_1 P i_2k_1 k_2}(p_{1}^{\prime 0}-p_{1}^{0})}
             {(p_{1}^{0}-\varepsilon_{k_1}+i0)
               (E-p_{1}^{0}-\varepsilon_{k_2}+i0)}
             \,.
   \label{g1}
\eeqn
Employing the identities
\beq
 \frac{1}{(p_{1}^{\prime 0}-\varepsilon_{P i_1}+i0)
       (E-p_{1}^{\prime 0}-\varepsilon_{P i_2}+i0)}
     = \frac{1}{E-E_{i}^{(0)}}
\left(\frac{1}{p_{1}^{\prime 0}-\varepsilon_{P i_1}+i0}
  +    \frac{1}{E-p_{1}^{\prime 0}-\varepsilon_{P i_2}+i0}
\right)\,,
\eeq
\beq
  \frac{1}{(p_{1}^{0}-\varepsilon_{k_1}+i0)
       (E-p_{1}^{0}-\varepsilon_{k_2}+i0)}
     = \frac{1}{E-E_{k}^{(0)}}
\left( \frac{1}{p_{1}^{0}-\varepsilon_{k_1}+i0}+
       \frac{1}{E-p_{1}^{0}-\varepsilon_{k_2}+i0}\right) \,,
     \label{ident}
\eeq
      we obtain
\beqn
     K_{ik}^{(1)}&=& F_i F_k
   \frac{1}{2\pi i}\oint_{\Gamma} dE
\frac{E}{(E-E_i^{(0)})(E-E_k^{(0)})}
  \Biggl[\left(\frac{i}{2\pi}\right)^{2}
 \int_{-\infty}^{\infty}dp_{1}^{0}\,dp_{1}^{\prime 0}\,
             \sum_{P}(-1)^{P}  \nonumber \\
 &&          \times \left( \frac{1}{p_{1}^{\prime 0}-\varepsilon_{P i_1}+i0}
              +   \frac{1}{E-p_{1}^{\prime 0}-\varepsilon_{P i_2}+i0}\right)
              \left( \frac{1}{p_{1}^{0}-\varepsilon_{k_1}+i0}+
           \frac{1}{E-p_{1}^{0}-\varepsilon_{k_2}+i0}\right) \nonumber \\
 &&       \times
I_{P i_1 P i_2 k_1 k_2}(p_{1}^{\prime 0}-p_{1}^{0}) \Biggr]\,.
   \label{K1}
\eeqn
The expression in the square brackets is an analytical function of $E$ inside
the contour $\Gamma$, if the photon mass $\mu$ is chosen properly (see
Refs.~\cite{shabaev:94:ttg2,shabaev:93}). Carrying out the $E$ integration by
Cauchy's theorem and taking into account that
\begin{equation}
         \left(\frac{i}{2 \pi}\right)\left(\frac{1}{x+i0}+
                      \frac{1}{-x+i0}\right)=\delta (x)\,,
    \label{del}
\end{equation}
we obtain
\beqn
K_{ik}^{(1)}&=& F_i F_k \left\{  \frac{i}{2\pi}
 \int_{-\infty}^{\infty}dp_{1}^{0}
             \sum_{P}(-1)^{P}
\frac{E_i^{(0)} I_{P i_1 P i_2 k_1 k_2}(\varepsilon_{Pi_1}-p_{1}^{0})
}{E_i^{(0)}-E_k^{(0)}} \right.
    \nonumber \\ &&\times
\left( \frac{1}{p_{1}^{0}-\varepsilon_{k_1}+i0}+
   \frac{1}{E_i^{(0)}-p_{1}^{0}-\varepsilon_{k_2}+i0}\right)
    \nonumber \\  &&
    + \frac{i}{2\pi}
 \int_{-\infty}^{\infty}dp_{1}^{\prime 0}
             \sum_{P}(-1)^{P}
\frac{E_k^{(0)}  I_{P i_1 P i_2 k_1 k_2}(p_{1}^{\prime 0}-\varepsilon_{k_1})
}{E_k^{(0)}-E_i^{(0)}}
 \nonumber \\ &&\times
 \left.
  \left( \frac{1}{p_{1}^{\prime
0}-\varepsilon_{P i_1}+i0}
              +   \frac{1}{E_{k}^{(0)}
-p_{1}^{\prime 0}-\varepsilon_{P i_2}+i0}\right)\right\}\,.
   \label{K1a}
\eeqn
In the same way we find
     \beqn
P_{ik}^{(1)}&=& F_i F_k  \left\{ \frac{i}{2\pi}
 \int_{-\infty}^{\infty}dp_{1}^{0}
             \sum_{P}(-1)^{P}
\frac{ I_{ P i_1 P i_2 k_1 k_2}(\varepsilon_{Pi_1}-p_{1}^{0})
}{E_i^{(0)}-E_k^{(0)}} \right.
  \nonumber \\ &&\times
  \left( \frac{1}{p_{1}^{0}-\varepsilon_{k_1}+i0}+
   \frac{1}{E_i^{(0)}-p_{1}^{0}-\varepsilon_{k_2}+i0}\right) \nonumber \\
 &&+
 \frac{i}{2\pi}
 \int_{-\infty}^{\infty}dp_{1}^{\prime 0}
             \sum_{P}(-1)^{P}
\frac{ I_{P i_1 P i_2 k_1 k_2}(p_{1}^{\prime 0}-\varepsilon_{k_1})
}{E_k^{(0)}-E_i^{(0)}}
 \nonumber \\ &&\times
   \left.
 \left( \frac{1}{p_{1}^{\prime 0}-\varepsilon_{P i_1}+i0}
              +   \frac{1}{E_{k}^{(0)}
-p_{1}^{\prime 0}-\varepsilon_{P i_2}+i0}\right)\right\}\,.
   \label{P1}
\eeqn
Substituting Eqs.~(\ref{K1a}), (\ref{P1}) into Eq.~(\ref{H1}), we get
\beqn
H_{ik}^{(1)}&=& F_i F_k \left\{ \frac{i}{4\pi}
 \int_{-\infty}^{\infty}dp_{1}^{0}\,
             \sum_{P}(-1)^{P}
 I_{ P i_1 P i_2 k_1 k_2}(\varepsilon_{Pi_1}-p_{1}^{0}) \right.
  \nonumber \\ &&\times
  \left( \frac{1}{p_{1}^{0}-\varepsilon_{k_1}+i0}+
   \frac{1}{E_i^{(0)}-p_{1}^{0}-\varepsilon_{k_2}+i0}\right) \nonumber \\
 &&+
\frac{i}{4\pi}
 \int_{-\infty}^{\infty}dp_{1}^{\prime 0}\,
             \sum_{P}(-1)^{P}
 I_{ P i_1 P i_2 k_1 k_2}(p_{1}^{\prime 0}-\varepsilon_{k_1})
    \nonumber \\ &&\times
 \left.
 \left( \frac{1}{p_{1}^{\prime 0}-\varepsilon_{P i_1}+i0}
              +   \frac{1}{E_{k}^{(0)}
-p_{1}^{\prime 0}-\varepsilon_{P i_2}+i0}\right)\right\}\,.
   \label{HH1}
\eeqn
Introducing the notations
 $\Delta_1=\varepsilon_{Pi_1}-\varepsilon_{k_1}$
and
 $\Delta_2=\varepsilon_{Pi_2}-\varepsilon_{k_2}$, we can rewrite Eq.~(\ref{HH1})
as follows,
\beqn
H_{ik}^{(1)}&=& F_i F_k
 \frac{i}{8\pi}
 \int_{-\infty}^{\infty}d\omega
             \sum_{P}(-1)^{P}
 I_{ P i_1 P i_2 k_1 k_2}(\omega)
 \left(
 \frac{1}{\omega+\Delta_1+i0}+
 \frac{1}{\Delta_2-\omega+i0} \right.
 \nonumber\\
&&+ \frac{1}{\omega+\Delta_2+i0}+
 \frac{1}{\Delta_1-\omega+i0}+
 \frac{1}{\omega-\Delta_1+i0}+
 \frac{1}{-\Delta_2-\omega+i0}
  \nonumber \\ && \left.
 +\frac{1}{\omega-\Delta_2+i0}+
 \frac{1}{-\Delta_1-\omega+i0}\right)\nonumber\\
&=&
F_i F_k  \frac{1}{4}
 \int_{-\infty}^{\infty}d\omega
             \sum_{P}(-1)^{P}
 I_{P i_1 P i_2k_1 k_2}(\omega)
  \nonumber \\ && \times
 \Bigl[\delta(\omega+\Delta_1)+\delta(\omega-\Delta_1)+
\delta(\omega+\Delta_2)+\delta(\omega-\Delta_2)\Bigr] \,.
\eeqn
Taking into account that $I(z)=I(-z)$, we finally obtain
\cite{mittleman:72,shabaev:93}
\beqn  \label{ope}
H_{ik}^{(1)}= F_i F_k \frac{1}{2}\sum_{P}(-1)^P [I_{Pi_1 Pi_2 k_1
k_2}(\Delta_1)+I_{Pi_1 Pi_2 k_1 k_2}(\Delta_2)]\,.
\eeqn

%%%%%%%%%%%%%%%%%%%%%%%%%%%%%%%%%%%%%%%%%%%%%%%%%%%%%%%%%%%%%%%%%%%%%%%%%%%%
\subsection{Two-photon exchange diagrams}

The set of two-photon exchange diagrams is shown in Fig.~\ref{fig:twophot}.
The first and the second graph are referred to as the ladder and the crossed
diagram, respectively. The derivation of the general expressions for the
two-photon exchange correction in the case of quasi-degenerate levels is
rather lengthy. However, it greatly resembles the corresponding derivation
for the one-photon exchange correction presented above, on one hand, and that
for the two-photon exchange diagram in case of a single level described in
detail in Ref.~\cite{shabaev:94:ttg1}, on the other hand. We thus present
only the final formulas for the two-photon exchange contributions to the
matrix elements of the operator $H^{(2)}$.

%%%%%%%%%%%%%%%%%%%%%%%%%%%%%%%%%%%%%%%%%
\subsubsection{The ladder diagram}

The contribution of the two-photon ladder diagram is conveniently divided
into the {\it irreducible} and the {\it reducible} part. The reducible
contribution is defined as a part in which the total intermediate energy of
the atom equals to $E_{1}^{(0)}$ or $E_{2}^{(0)}$ and the irreducible part is
the remainder. The operator $H^{(2)}$ is defined by Eq.~(\ref{H2}). The first
three terms in the right-hand side of this equation contribute both to the
irreducible and to the reducible part. As to the others, it is natural to
ascribe them to the reducible part.

The contribution of the irreducible part of $H^{(2)}_{ik}$ is defined
as the self-adjoint part of the following matrix
\beq
H_{ik}^{\rm lad,\, ir} = [K^{(2,{\rm ir})}- (1/2)P^{(2,{\rm
ir})}K^{(0)}-(1/2)K^{(0)}P^{(2,{\rm ir})}]_{ik}\,.
\eeq
The result is
\beqn  \label{Hir}
H_{ik}^{{\rm lad,\, ir}} &=&
    F_i F_k\Biggl\{\frac14\left[
    S_{ik}(E_i^{(0)},0,0)+S_{ik}(E_i^{(0)},0,\Delta)
    +S_{ik}(E_k^{(0)},0,0)+S_{ik}(E_k^{(0)},-\Delta,0) \right]
  \nonumber\\    &&
    +\frac{i}{4\pi}\,{\rm v.p.}\!\! \int_{-\infty}^{\infty} dx\,\frac{1}{x}
        \left[S_{ik}(E_i^{(0)},0,x)-S_{ik}(E_i^{(0)},0,x+\Delta)
    \right.
 \nonumber\\ &&
    \left.
 +S_{ik}(E_k^{(0)},x,0)-S_{ik}(E_k^{(0)},x-\Delta,0)\right]\Biggr\}\,,
\eeqn
where $\Delta = E_i^{(0)}-E_k^{(0)}$ and the matrix elements $S_{ik}$ are
defined by
\beqn  \label{S}
S_{ik}(E,x,y)&=&\sum_{P}(-1)^P\frac{i}{2\pi}
  \nonumber \\ && \!\!\!\!\!\!\!  \times
\int_{-\infty}^{\infty} d\omega\sum_{n_1
n_2}^{E_n^{(0)}\not=E_1^{(0)}\!,E_2^{(0)}}\frac{ I_{Pi_1 Pi_2 n_1
n_2}(\varepsilon_{Pi_1}-\omega+x)\, I_{n_1 n_2 k_1
k_2}(\varepsilon_{k_1}-\omega+y)} {[\omega-\varepsilon_{n_1}(1-i0)]
[E-\omega-\varepsilon_{n_2}(1-i0)]}\,.
\eeqn
The summation here runs over all $n_1$ and $n_2$ for which
$E_n^{(0)}\not=E_1^{(0)}, E_2^{(0)}$, where $E_{n}^{(0)}\equiv
\varepsilon_{n_1}+\varepsilon_{n_2}$ is the total intermediate energy of the
atom. The sign "v.p." in front of the integral in Eq.~(\ref{Hir}) denotes
that the principal value of the integral (over $x$) must be taken.

We note that the part containing the integral over $x$ in Eq.~(\ref{Hir})
vanishes identically in case of diagonal matrix elements ($i = k$). It
neither appears for single levels \cite{shabaev:94:ttg1}. In case of
off-diagonal matrix elements ($i \ne k$), the contribution of this part is of
order $\alpha^2 \Delta$ and it vanishes when $(E^{(0)}_i-E^{(0)}_k)\to 0$. As
shown in Ref.~\cite{shabaev:02:rep}, such terms contribute to the next order
of perturbation theory and can, therefore, be disregarded in the present
consideration. Expression (\ref{Hir}) can be simplified even further by
taking into account that
\beq    \label{as:1}
E_i^{(0)} = \overline{E}^{(0)} + O(\Delta)\,, \ \ \ E_k^{(0)} =
\overline{E}^{(0)} + O(\Delta)\,,
\eeq
 where $\overline{E}^{(0)} =
(E_i^{(0)}+E_k^{(0)})/2$. We thus write $H_{ik}^{{\rm lad,\, ir}}$ simply as
\beqn   \label{Hirsimple}
 H_{ik}^{{\rm lad,\, ir}} &=& F_i F_k\,S_{ik}(\overline{E}^{(0)},0,0)+
        O(\alpha^2\Delta)
  \nonumber \\
  &=&
  F_i F_k\, \sum_{P}(-1)^P\frac{i}{2\pi}
%  \nonumber \\ && \!\!\!\!\!\!\!  \times
\int_{-\infty}^{\infty} d\omega
   \nonumber \\ && \times
\sum_{n_1 n_2}^{E_n^{(0)}\not=E_1^{(0)}\!,E_2^{(0)}}\frac{ I_{Pi_1 Pi_2 n_1
n_2}(\varepsilon_{Pi_1}-\omega)\, I_{n_1 n_2 k_1
k_2}(\varepsilon_{k_1}-\omega)} {[\omega-\varepsilon_{n_1}(1-i0)]
[\overline{E}^{(0)}-\omega-\varepsilon_{n_2}(1-i0)]} + O(\alpha^2\Delta)\,.
\eeqn

The reducible contribution is induced by the self-adjoint part of the
following operator
\beqn
H^{{\rm lad,\, red}}=H^{{\rm lad,\, red},a}+H^{{\rm lad,\, red},b}\,,
\eeqn
where
\beqn  \label{reda}
H^{ {\rm lad,\, red}, a}= K^{(2,{\rm\, red})}- \frac12P^{(2,{\rm
   \, red})}K^{(0)}-\frac12K^{(0)}P^{(2,{\rm\, red})}\,
\eeqn
and
\beqn  \label{redb}
H^{{\rm lad,\, red},b}&=&-\frac 12 P^{(1)}K^{(1)}-\frac 12 K^{(1)}P^{(1)}
    +\frac 38 P^{(1)}P^{(1)}K^{(0)}
    \nonumber \\&&
    +\frac 38 K^{(0)}P^{(1)}P^{(1)}
    +\frac 14 P^{(1)}K^{(0)}P^{(1)}\,.
\eeqn

The result for the first part reads
\beqn \label{Hred}
     H_{ik}^{{\rm lad,\, red},a}&=& F_i F_k\Biggl\{
-\frac12 \left[A_{ik}(0)+A_{ik}(\Delta)+B_{ik}(0)
+B_{ik}(-\Delta)+C_{ik}\right]
    \nonumber\\ &&
    -\frac14 \left[D_{ik}(E_i^{(0)},0,0)+D_{ik}(E_i^{(0)},0,\Delta)+
D_{ik}(E_k^{(0)},0,0)+D_{ik}(E_k^{(0)},-\Delta,0)\right]
    \nonumber\\ &&
    -\frac{i}{4\pi}\,{\rm v.p.}\!\!\int_{-\infty}^{\infty}
dx\,\frac{1}{x} \left[D_{ik}(E_i^{(0)},0,x)-D_{ik}(E_i^{(0)},0,x+\Delta)
    \right.
    \nonumber\\&&
    \left.
    +D_{ik}(E_k^{(0)},x,0)-D_{ik}(E_k^{(0)},x-\Delta,0)\right]\Biggr\}\,,
\eeqn
where
\beqn  \label{Hreda}
A_{ik}(x)&=&\sum_{P}(-1)^P\frac{i}{2\pi} \sum_{n_1
n_2}^{E_n^{(0)}=E_1^{(0)},E_2^{(0)}}\nonumber \\ &&\times
\int_{-\infty}^{\infty} d\omega\,\frac{ I_{Pi_1 Pi_2 n_1
n_2}(\omega-\varepsilon_{n_1})\, I_{n_1 n_2 k_1
k_2}(\varepsilon_{k_1}-\varepsilon_{n_1})}
{(\omega-\varepsilon_{Pi_1}+E_i^{(0)}-E_n^{(0)}-i0)
(\omega-\varepsilon_{Pi_1}+x-i0) }\,,
\eeqn
\beqn
B_{ik}(x)&=&\sum_{P}(-1)^P\frac{i}{2\pi} \sum_{n_1
n_2}^{E_n^{(0)}=E_1^{(0)},E_2^{(0)}}\nonumber \\ &&\times
\int_{-\infty}^{\infty} d\omega\,\frac{ I_{Pi_1 Pi_2 n_1
n_2}(\varepsilon_{Pi_1}-\varepsilon_{n_1}) \,I_{n_1 n_2 k_1
k_2}(\omega-\varepsilon_{n_1})}
{(\omega-\varepsilon_{k_1}+E_k^{(0)}-E_n^{(0)}-i0)
(\omega-\varepsilon_{k_1}+x-i0) }\,,
\eeqn
\beqn
 C_{ik}&=&\sum_{P}(-1)^P \sum_{n_1
n_2}^{E_n^{(0)}=E_1^{(0)},E_2^{(0)}}
(E_i^{(0)}+E_k^{(0)}-2E_n^{(0)})\nonumber\\
&&\times\frac{i}{2\pi}\int_{-\infty}^{\infty} d\omega'\,\frac{ I_{Pi_1 Pi_2
n_1 n_2}(\omega'-\varepsilon_{n_1})} {(\omega'-\varepsilon_{Pi_1}-i0)
(\omega'-\varepsilon_{Pi_1}+E_i^{(0)}-E_n^{(0)}-i0) }\nonumber\\ &&\times
\frac{i}{2\pi}\int_{-\infty}^{\infty} d\omega\,\frac{ I_{n_1 n_2 k_1
k_2}(\omega-\varepsilon_{n_1})} {(\omega-\varepsilon_{k_1}-i0)
(\omega-\varepsilon_{k_1}+E_k^{(0)}-E_n^{(0)}-i0)} \,,
\eeqn
\beqn     \label{Hredd}
D_{ik}(E,x,y)&=&\sum_{P}(-1)^P\frac{i}{2\pi}\nonumber \sum_{n_1
n_2}^{E_n^{(0)}=E_1^{(0)}, E_2^{(0)}}\\ &&\times \int_{-\infty}^{\infty}
d\omega\,\frac{ I_{Pi_1 Pi_2 n_1 n_2}(\varepsilon_{Pi_1}-\omega+x) \,I_{n_1
n_2 k_1 k_2}(\varepsilon_{k_1}-\omega+y)} {(\omega-\varepsilon_{n_1}-i0)
(\omega+\varepsilon_{n_2}-E-i0) }\,.\label{Dred}
\eeqn
The part containing the integral over $x$
in Eq.~(\ref{Hred}) represents a contribution of
order $\alpha^2\Delta$. Again, we
regard this contribution as belonging to the next order of perturbation
theory and disregard it in the present investigation.

The second part of the reducible contribution is given by the matrix element
of the operator (\ref{redb}). The result is obtained by taking into account
that
\beqn  \label{Kmat_0}
K_{ik}^{(0)}=E_i^{(0)}\delta_{ik}\,,\;\;\;\;\;\;\;\;\;
P_{ik}^{(0)}=\delta_{ik}\,,
\eeqn
\beqn
K_{ik}^{(1)}&=& F_i F_k \sum_{P}(-1)^{P} \left\{ \frac{1}{2}\,\left[I_{ P i_1
P i_2 k_1 k_2} (\Delta_1)+I_{ P i_1 P i_2 k_1 k_2}(\Delta_2)\right]
    \right.
    \nonumber\\ &&
-\frac{(E_i^{(0)}+E_k^{(0)})}{2}
 \frac{i}{2\pi}
 \int_{-\infty}^{\infty}d\omega\,
I_{ P i_1 P i_2 k_1 k_2} (\omega)
    \nonumber\\ &&\times
        \left.
\left[ \frac{1}{(\omega+\Delta_1-i0) (\omega-\Delta_2-i0)}
+\frac{1}{(\omega+\Delta_2-i0) (\omega-\Delta_1-i0)}\right]\right\}\,,
\eeqn
and
\beqn
P_{ik}^{(1)}&=& -F_i F_k \sum_{P}(-1)^{P}
 \frac{i}{2\pi}
 \int_{-\infty}^{\infty}d\omega\,
I_{ P i_1 P i_2 k_1 k_2} (\omega)\nonumber\\ &&\times \left[ \frac{1}{
(\omega+\Delta_1-i0) (\omega-\Delta_2-i0)} +\frac{1}{ (\omega+\Delta_2-i0)
(\omega-\Delta_1-i0)}\right]\,.\label{Pmat_1}
\eeqn

The total result for the reducible part can be simplified by using
Eq.~(\ref{as:1}) and disregarding terms that contribute to the next order of
perturbation theory. One can show that in this case the $A$'s, $B$'s, and
$C$'s in Eq.~(\ref{Hred}) are cancelled completely by the $H^{{\rm lad,
red},b}$ term. The result is just
\beqn
H_{ik}^{{\rm lad,\, red}}&=& -F_i F_k\, D_{ik}(\overline{E}^{(0)},0,0)+
        O(\alpha^2\Delta)
%    \nonumber \\ &=&
= -F_i F_k\,\sum_{P}(-1)^P \sum_{n_1 n_2}^{E_n^{(0)}=E_1^{(0)}, E_2^{(0)}}
    \nonumber\\ &&\times
 \frac{i}{2\pi} \int_{-\infty}^{\infty} d\omega\,\frac{ I_{Pi_1 Pi_2 n_1
n_2}(\varepsilon_{Pi_1}-\omega) \,I_{n_1 n_2 k_1
k_2}(\varepsilon_{k_1}-\omega)} {(\omega-\varepsilon_{n_1}-i0)
(\omega+\varepsilon_{n_2}-\overline{E}^{(0)}-i0) } + O(\alpha^2\Delta)\,.
\eeqn

%%%%%%%%%%%%%%%%%%%%%%%%%%%%%%%%%%%%
\subsubsection{The crossed diagram}

The contribution of the crossed diagram is induced by the self-adjoint part
of the following operator
\beq
H^{\rm cr}= K^{(2)}- (1/2)P^{(2)}K^{(0)}-(1/2)K^{(0)}P^{(2)}\,.
\eeq
The corresponding result reads
\beqn  \label{Hcr}
H_{ik}^{\rm cr} &=& F_i F_k\Biggl\{\frac14\left[
T_{ik}(E_i^{(0)},0,0)+T_{ik}(E_i^{(0)},0,\Delta)
+T_{ik}(E_k^{(0)},0,0)+T_{ik}(E_k^{(0)},-\Delta,0)\right]
    \nonumber\\&&
+\frac{i}{4\pi}\,{\rm v.p.}\!\!\int_{-\infty}^{\infty} dx\,\frac{1}{x}
\left[T_{ik}(E_i^{(0)},0,x)-T_{ik}(E_i^{(0)},0,x+\Delta)
    \right.
    \nonumber\\&&
    \left.
    +T_{ik}(E_k^{(0)},x,0)-T_{ik}(E_k^{(0)},x-\Delta,0)\right]\Biggr\}\,,
\eeqn
where
\beqn  \label{T}
T_{ik}(E,x,y)&=&\sum_{P}(-1)^P \sum_{n_1 n_2} \frac{i}{2\pi}
    \nonumber\\&&
    \!\!\!\!\!\!\!\!\!\times
    \int_{-\infty}^{\infty} d\omega\,\frac{ I_{Pi_1 n_2 n_1
k_2}(\varepsilon_{Pi_1}-\omega+x)\, I_{n_1 Pi_2 k_1
n_2}(\varepsilon_{k_1}-\omega+y)} {[\omega-\varepsilon_{n_1}(1-i0)]
[E-\varepsilon_{Pi_1}-\varepsilon_{k_1}-x-y+\omega -\varepsilon_{n_2}
(1-i0)]}\,.
\eeqn
The expression (\ref{Hcr}) can be simplified in the same way as the previous
contributions, with the result
\beqn
H_{ik}^{\rm cr}&=& F_i F_k\,
    T_{ik}(\overline{E}^{(0)},0,0)+  O(\alpha^2\Delta)
%    \nonumber\\  &=&
=  F_i F_k\,\sum_{P}(-1)^P \sum_{n_1 n_2} \frac{i}{2\pi}
    \nonumber\\&&
    \!\!\!\!\!\!\!\!\!\times
    \int_{-\infty}^{\infty} d\omega\,\frac{ I_{Pi_1 n_2 n_1
k_2}(\varepsilon_{Pi_1}-\omega)\, I_{n_1 Pi_2 k_1
n_2}(\varepsilon_{k_1}-\omega)} {[\omega-\varepsilon_{n_1}(1-i0)]
[\overline{E}^{(0)}-\varepsilon_{Pi_1}-\varepsilon_{k_1}+\omega
-\varepsilon_{n_2} (1-i0)]}
%    \nonumber \\ &&
    +  O(\alpha^2\Delta)\,.
\eeqn

%%%%%%%%%%%%%%%%%%%%%%%%%%%%%%%%%%%%%%%%%%%%%%%%%%%%%%%%%%%%%%%%%%%%%%%
\subsection{Screened self-energy correction}

The set of Feynman diagrams representing the screened self-energy correction
is shown in Fig.~\ref{fig:sevpscr}. Formal expressions for this correction in
case of quasi-degenerate states were obtained previously in
Ref.~\cite{bigot:01} by the TTGF method. Here we present only the final
expressions for this correction.

The contribution of the vertex diagrams is given by
\beqn
H_{ik}^{\rm ver} &=& F_iF_k\,\sum_P (-1)^P
    \frac{i}{2\pi} \int_{-\infty}^{\infty}d\omega\,\sum_{n_1n_2}
    \left\{
        \frac{I_{ n_1Pi_2n_2k_2}(\Delta_1)
            \, I_{Pi_1n_2n_1k_1}(\omega)}
            {[\eps_{Pi_1}-\omega-\eps_{n_1}(1-i0)]
            [\eps_{k_1}-\omega-\eps_{n_2}(1-i0)]}
 \right.
    \nonumber \\ &&
 \left.   +
        \frac{I_{ Pi_1n_1k_1n_2}(\Delta_2)
            \, I_{Pi_2n_2n_1k_2}(\omega)}
            {[\eps_{Pi_2}-\omega-\eps_{n_1}(1-i0)]
            [\eps_{k_2}-\omega-\eps_{n_2}(1-i0)]}
                \right\} + O(\alpha^2\Delta)\,,
\eeqn
where $\Delta_1=\varepsilon_{Pi_1}-\varepsilon_{k_1}$ and
 $\Delta_2=\varepsilon_{Pi_2}-\varepsilon_{k_2}$.

The contribution of the remaining diagrams is conveniently separated into the
irreducible and reducible parts. The irreducible contribution is given by
\beqn
H_{ik}^{\rm se, ir} &=& F_iF_k\,\sum_P (-1)^P \left\{
    \sum_{n\ne k_1} \frac{I_{Pi_1Pi_2nk_2}(\Delta_1)}{\eps_{k_1}-\eps_n}
            \lbr n |\Sigma(\eps_{k_1})|k_1\rbr
            \right.
    \nonumber \\ &&
    +\sum_{n\ne k_2} \frac{I_{Pi_1Pi_2k_1n}(\Delta_2)}{\eps_{k_2}-\eps_n}
            \lbr n |\Sigma(\eps_{k_2})|k_2\rbr
    \nonumber \\ &&
    +\sum_{n\ne Pi_1}  \lbr Pi_1 |\Sigma(\eps_{Pi_1})|n\rbr
            \frac{I_{nPi_2k_1k_2}(\Delta_1)}{\eps_{Pi_1}-\eps_n}
    \nonumber \\ &&
        \left.
    +\sum_{n\ne Pi_2}  \lbr Pi_2 |\Sigma(\eps_{Pi_2})|n\rbr
            \frac{I_{Pi_1nk_1k_2}(\Delta_2)}{\eps_{Pi_2}-\eps_n}
            \right\}  + O(\alpha^2\Delta) \,,
\eeqn
where $\Sigma(\eps)$ is the self-energy operator defined by its matrix
elements,
\begin{equation} %\label{}
\langle a|\Sigma (\varepsilon )|b \rangle = \frac{i}{2\pi}
\int_{-\infty}^{\infty} d\omega \, \sum_n
 \frac{\langle an|I(\omega )|nb \rangle}{\varepsilon -\omega
   - {\varepsilon}_n (1-i0)} \,.
\end{equation}

The result for the reducible contribution reads
\beqn
H_{ik}^{\rm se, red} &=& F_iF_k\,\frac12 \sum_P (-1)^P \Bigl\{
    I_{Pi_1Pi_2k_1k_2}(\Delta_1)
        \Bigl[ \lbr Pi_1|\Sigma^{\prime}(\eps_{Pi_1})|Pi_1\rbr
            + \lbr k_1|\Sigma^{\prime}(\eps_{k_1})|k_1\rbr \Bigr]
     \nonumber \\ &&
     + I_{Pi_1Pi_2k_1k_2}(\Delta_2)
        \Bigl[ \lbr Pi_2|\Sigma^{\prime}(\eps_{Pi_2})|Pi_2\rbr
            + \lbr k_2|\Sigma^{\prime}(\eps_{k_2})|k_2\rbr \Bigr]
     \nonumber \\ &&
    +I^{\prime}_{Pi_1Pi_2k_1k_2}(\Delta_1)
        \Bigl[ \lbr Pi_1|\Sigma(\eps_{Pi_1})|Pi_1\rbr
            - \lbr k_1|\Sigma(\eps_{k_1})|k_1\rbr \Bigr]
     \nonumber \\ &&
     + I^{\prime}_{Pi_1Pi_2k_1k_2}(\Delta_2)
        \Bigl[ \lbr Pi_2|\Sigma(\eps_{Pi_2})|Pi_2\rbr
            - \lbr k_2|\Sigma(\eps_{k_2})|k_2\rbr \Bigr]
            \Bigr\} + O(\alpha^2\Delta)\,,
\eeqn
where $I^{\prime}(\omega) \equiv \partial I(\omega)/\partial \omega$, and
$\Sigma^{\prime}(\omega) \equiv \partial \Sigma(\omega)/\partial \omega$.

%%%%%%%%%%%%%%%%%%%%%%%%%%%%%%%%%%%%%%%%%%%%%%%%%%%%%%%%%%%%%%%%%%%%%%%
\subsection{Screened vacuum-polarization correction}

The derivation of formal expressions for the screened vacuum-polarization
correction in case of quasi-degenerate states was described in our previous
work \cite{artemyev:00:pra}. For completeness, we present here the final
expressions for this correction; the corresponding set of Feynman diagrams is
shown in Fig.~\ref{fig:sevpscr}.

The expression for the contribution of the diagram with the
vacuum-polarization loop inserted into the photon propagator can be obtained
from the formula for the one-photon exchange (\ref{ope}) by replacing the
operator of the electron-electron interaction $I(\eps)$ by the modified
interaction,
\beqn  \label{vps1}
U_{\rm VP}^{\rm ph}(\eps,\bfx,\bfy)&=&\frac{\alpha^2}{2\pi i}\int_{-\infty}^\infty
d \omega\, \int d\bfz_1\,  d\bfz_2\,
\frac{\alpha_{\mu}\exp(i|\eps||\bfx-\bfz_1|)} {|\bfx-\bfz_1|}\,
\frac{\alpha_{\nu}\exp(i|\eps||\bfy-\bfz_2|)} {|\bfy-\bfz_2|}\,
    \nonumber \\&&
    \times \rm{Tr}\left[\alpha^\mu G(\omega-\eps /2,\bfz_1,\bfz_2)
    \,\alpha^\nu\,
G(\omega+\eps /2,\bfz_2,\bfz_1)\right]\,,
\eeqn
where $G(\omega,\bfx,\bfy)=\sum_n \psi_n(\bfx)
\psi_n^\dag(\bfy)/[\omega-\eps_n(1-i0)]$ is the Dirac-Coulomb Green function.
The corresponding contribution to $H_{ik}^{(2)}$ is
\beqn
H_{ik}^{\rm vp,\,ph} &=& F_i F_k\, \frac{1}{2}\sum_{P}(-1)^P
  \left[ \lbr Pi_1 Pi_2|U_{\rm VP}^{\rm ph}(\Delta_1)| k_1 k_2\rbr
+ \lbr Pi_1 Pi_2|U_{\rm VP}^{\rm ph}(\Delta_2)| k_1 k_2\rbr \right]\,,
\eeqn
where $\Delta_1=\varepsilon_{Pi_1}-\varepsilon_{k_1}$ and
$\Delta_2=\varepsilon_{Pi_2}-\varepsilon_{k_2}$.

To the order under consideration, expressions for the remaining diagrams can
be obtained from the one-photon exchange correction by perturbing the wave
functions and the binding energies by an additional vacuum-polarization
interaction. The result is
\beqn  \label{vps2}
H_{ik}^{\rm vp, wf}+
H_{ik}^{\rm vp, be}&=&F_i F_k\, \frac 12
        \sum_{P}(-1)^P \Bigl[
\lbr \delta Pi_1 Pi_2 | \left[I(\Delta_1)+I(\Delta_2)\right]| k_1 k_2 \rbr
    \nonumber \\ &&
+\lbr Pi_1 \delta Pi_2| \left[I(\Delta_1)+I(\Delta_2)\right]|  k_1 k_2\rbr
    \nonumber \\ &&
+\lbr Pi_1 Pi_2 |\left[I(\Delta_1)+I(\Delta_2)\right]| \delta k_1 k_2\rbr
    \nonumber \\ &&
+\lbr Pi_1 Pi_2 |\left[I(\Delta_1)+I(\Delta_2)\right]| k_1 \delta k_2\rbr
         \nonumber \\ &&
+(\delta\eps_{Pi_1}-\delta\eps_{k_1})\,
     \lbr Pi_1 Pi_2 |I^{\prime}(\Delta_1)| k_1 k_2 \rbr
         \nonumber \\ &&
+(\delta\eps_{Pi_2}-\delta\eps_{k_2})\,
     \lbr Pi_1 Pi_2 |I^{\prime}(\Delta_2)| k_1 k_2 \rbr
\Bigr]\,.
\eeqn
where $\delta i$ and $\delta k$ refer to the first-order corrections to the
corresponding wave function,
\beq
|\delta i\rbr = \sum_{n}^{\eps_n \ne \eps_i}\frac{|n\ra\la
  n|U_{\rm VP}|i\ra}{\eps_i-\eps_n}\,,
\eeq
$\delta \eps_i$ is the correction to the energy,
$
\delta\eps_i = \la i|U_{\rm VP}|i\ra\,,
$
and
\beqn
U_{\rm VP}(\bfx)=\frac{\alpha}{2\pi i}\int_{-\infty}^\infty d \omega \int
d\bfy \frac{1}{|\bfx-\bfy|}\rm{Tr}\left[G(\omega,\bfy,\bfy)\right]
\eeqn
is the vacuum-polarization potential.

As discussed previously in Ref.~\cite{shabaev:02:rep}, a direct derivation
based on the TTGF method yields a result that differs from Eq.~(\ref{vps2})
by terms of order $(\alpha^2\Delta)$, which can be disregarded as long as we
are not interested in higher orders of perturbation theory (see
Ref.~\cite{shabaev:02:rep} for a detailed discussion).

%%%%%%%%%%%%%%%%%%%%%%%%%%%%%%%%%%%%%%%%%%%%%%%%%%%%%%%%%%%%%%%%%%%%%%%%%%%%%%%%%
%
%
%
%%%%%%%%%%%%%%%%%%%%%%%%%%%%%%%%%%%%%%%%%%%%%%%%%%%%%%%%%%%%%%%%%%%%%%%%%%%%%%%%%
\section{Numerical evaluation and results}
\label{sec:twoelQED}

An important difference of the present investigation from the previous
studies of QED effects in high-$Z$ ions is that it involves QED corrections
for {\em quasi-degenerate} configurations, namely $(1s2p_{1/2})_1$ and
$(1s2p_{3/2})_1$. While the derivation of basic expressions in this case is
more difficult than for a single state, the final expressions for the
diagonal matrix elements turn out to be very similar to those for the
single-level case. We can, therefore, adopt a code developed for single-level
calculations for the diagonal matrix elements of the operator $H$. For an
evaluation of the off-diagonal matrix elements, a generalization of the code
is needed.

The numerical procedure employed in the present calculation of the two-photon
exchange correction is based on that presented in detail in our previous
investigations for Li-like ions \cite{yerokhin:00,artemyev:03}. Apart from
the angular reduction that is performed by using the standard
angular-momentum technique, the evaluation is rather similar to that for
Li-like ions. The calculation was carried out employing the Fermi model for
the nuclear-charge distribution, with the nuclear charge radii specified in
Section~\ref{sec:compilation}. The numerical uncertainty of the results is
expected to be $1\times10^{-4}$~eV in all cases except for the off-diagonal
matrix element, for which the uncertainty is $1\times10^{-4}$~eV for $Z\le
50$, $2\times10^{-4}$~eV for $Z\le 80$, and $4\times10^{-4}$~eV otherwise. As
a check of the numerical procedure, we performed the evaluation in two
different gauges, the Feynman and the Coulomb ones. The two-photon exchange
corrections (for mixing configurations, individual matrix elements) were
found to be gauge invariant well within the uncertainty specified.

The results of our numerical calculation of the two-photon exchange
correction for $n=1$ and $n=2$ states of He-like ions are presented in
Table~\ref{tab:2phot}.  The values listed represent corrections to the energy
in case of single levels and contributions to the matrix elements $H_{ik}$
for the quasi-degenerate states. The energy levels for the $(1s2p_{1/2})_1$
and $(1s2p_{3/2})_1$ states are obtained by diagonalizing the $2\times 2$
matrix $H$ containing all relevant corrections. In Table~\ref{tab:2phot}, we
present also a comparison of our numerical values with the results of the
previous calculations of this correction for various states of He-like ions
\cite{blundell:93:b,mohr:00:pra,asen:02,andreev:01,andreev:03,andreev:04}.
The comparison indicates that calculations by different groups are generally
in agreement with each other. However, there exist also certain deviations
between different calculations, notably with those by Andreev \etal
\cite{andreev:01,andreev:03}. Regarding the comparison of the present results
and the ones of Ref.~\cite{andreev:04} for the mixing states, we would like
to stress that, generally speaking, results of different methods for
individual matrix elements could be different, since the matrix $H$ can
differ by a unitary transformation. We observe, however, that in our case the
results for the individual matrix elements agree with those of
Ref.~\cite{andreev:04} approximately at the same level as for the single
states.

The calculation of the screened self-energy correction for $n=2$ states of
He-like ions resembles that for Li-like ions described in our previous work
\cite{yerokhin:99:sescr}. A more difficult angular structure of the
initial-state wave functions for He-like ions makes final expressions more
lengthy and their numerical evaluation more time consuming. Significant
complications appear in performing angular integrations in momentum space for
the vertex part with free-electron propagators. To handle them, we developed
a generalization of the angular-integration procedure described in
Ref.~\cite{yerokhin:99:sescr} to arbitrary states, using our experience in
calculating similar angular integrals for the two-loop self-energy diagrams
\cite{yerokhin:03:epjd}. The actual calculation was carried out employing the
spherical-shell model for the nuclear-charge distribution. Our numerical
results for the screened self-energy correction for $n=1$ and $n=2$ states of
He-like ions are presented in Table~\ref{tab:sescr} in terms of the
dimensionless function $F(\aZ)$ defined as
\beq    \label{FaZ}
\Delta E = \alpha^2 (\aZ)^3F(\aZ)\,.
\eeq
The values listed in the table represent corrections to the energy in case of
single levels and contributions to the matrix elements $H_{ik}$ for the
quasi-degenerate states.

In case of the ground state of He-like ions, the self-energy correction was
evaluated previously by Persson \etal \cite{persson:96:2el}, by us
\cite{yerokhin:97:pla}, and by Sunnergren \cite{sunnergren:98:phd}. In the
present work, we recalculated this correction using the new code and found an
excellent agreement with our previous results and with those by Sunnergren. A
small deviation of the present result for $Z=90$ from the old one is due to a
more recent value for the nuclear charge radius used in this work.

We note that the values presented in Table~\ref{tab:sescr} for $n=2$ states
of He-like ions can also be used for determining the screened self-energy
correction due to the interaction of the valence electron and the $(1s)^2$
shell in {\em Li-like} ions. Indeed, by using elementary angular-summation
rules, we obtain
\beq    \label{Lilike}
(2j_v+1)\, \Delta E^{\rm Li}_v = \sum_J (2J+1)\, \Delta E_{v,\, J}^{\rm
He}\,,
\eeq
where $\Delta E^{\rm Li}_v$ denotes the screened self-energy correction in a
Li-like ion due to the interaction of the electron in the state $v$ and the
$(1s)^2$ shell, $\Delta E_{v,\, J}^{\rm He}$ is the screened self-energy
correction in a He-like ion for the $(1s\,v)_J$ configuration (in case of
mixing configurations, a diagonal matrix element should be taken), and $j_v$
is the total angular momentum of the $v$ electron. By employing the identity
(\ref{Lilike}), we check that our numerical results for He-like ions are in a
very good agreement with our previous calculations for Li-like ions
\cite{yerokhin:99:sescr}.

Our calculations of the screened self-energy and two-photon exchange
corrections, combined with the results for the screened vacuum-polarization
from Ref.~\cite{artemyev:00:pra} (with the off-diagonal matrix elements
corrected in this paper, see below), complete the evaluation of the QED
correction to first order in $1/Z$ and to all orders in $\aZ$ for $n=2$
states of He-like ions. As is known, the $\aZ$ expansion of two-electron QED
effects starts with $\alpha^2(\aZ)^3$. The two-photon exchange correction
contains also contributions of previous orders in $\aZ$ that can be derived
from the Breit equation. We separate the ``pure" QED part of the two-photon
exchange contribution ($\Delta E_{\rm 2ph}^{\rm QED}$) as
\beq    \label{2photQED}
\Delta E_{\rm 2ph} = \alpha^2 [ a_0 + (\aZ)^2 a_2] + \Delta E_{\rm
            2ph}^{\rm QED}\,,
\eeq
where $\Delta E_{\rm 2ph}$ is the total two-photon exchange correction and
$\Delta E_{\rm 2ph}^{\rm QED}$ contributes to order $\alpha^2 (\aZ)^3$ and
higher. In order to extract numerical values for $\Delta E_{\rm 2ph}^{\rm
QED}$ from our results for $\Delta E_{\rm 2ph}$ without losses in accuracy,
accurate values for the coefficients $a_0$ and $a_2$ are needed. We calculate
them by fitting our results for the two-photon exchange correction obtained
within many-body perturbation theory. A large number of fitting points and
inclusion of fraction values for the nuclear charge number (up to $Z=0.1$)
allowed us to achieve better accuracy than in previous calculations of
similar coefficients (e.g.,
Refs.~\cite{sanders:69,aashamar:70,drake:CJP:88}). The numerical results for
the coefficients $a_0$ and $a_2$ for all states under consideration are
tabulated in the second and in the third column of Table~\ref{tab:coef},
respectively.

In Table~\ref{tab:twoelQED} we collect all two-electron QED contributions for
$n=1$ and $n=2$ states of He-like ions. The screened self-energy and
two-photon exchange corrections are calculated in the present work; in the
table they are labeled as ``Scr.SE" and ``2-ph.exch.", respectively.  The
screened vacuum-polarization correction was first evaluated in our previous
investigation \cite{artemyev:00:pra}. In the present work, we correct an
error made in Ref.~\cite{artemyev:00:pra} for the off-diagonal matrix element
and extend our calculation to the region $10 < Z < 20$. Numerical values for
the screened vacuum-polarization correction are listed in
Table~\ref{tab:twoelQED} under entry ``Scr.VP".

Our results for the two-electron QED correction calculated to all orders in
$\aZ$ can be compared with the results obtained within the $\aZ$ expansion,
which reads \cite{araki:57,sucher:58}
\beq    \label{2elQED}
\Delta E_{\rm 2el}^{\rm QED} = \alpha^2 (\aZ)^3 \biggl[ a_{31}\ln \aZ+
a_{30}+ (\aZ) \,G^{\rm h.o.}_{\rm 2el}(\aZ) \biggr]\,,
\eeq
where the function $G^{\rm h.o.}_{\rm 2el}(\aZ)$ is the higher-order
remainder that is not known analytically at present.  We obtain numerical
values for the coefficients $a_{31}$ and $a_{30}$ by using formulas from
Ref.~\cite{araki:57} and numerical results for the two-electron Bethe
logarithms \cite{goldman:84} and for the $1/Z$-expansion coefficients of
expectation values of various operators \cite{drake:CJP:88,drake:82:nimb}.
The only coefficient whose numerical value was not available in the
literature was the anomalous-magnetic moment correction for the off-diagonal
matrix element. This is the first-order $1/Z$-expansion term of the matrix
element of the operator
$\alpha/\pi(H_3^{\prime\prime\prime}+H_5^{\prime\prime\prime})$ (see
Eqs.~(27) and (28) of Ref.~\cite{drake:82:nimb}). The result of our
calculation of this correction (denoted in Ref.~\cite{drake:CJP:88} as
$\Delta E_{\rm anom}$) for the off-diagonal term in the LS coupling reads
\beq    \label{anom}
\Delta E_{\rm anom}^{LS}(\mbox{\rm offdiag}) = \alpha^2(\aZ)^3\,
        0.010110\,.
\eeq
Numerical values for the coefficients $a_{31}$ and $a_{30}$ for all states
under consideration are listed in the third and in the fourth columns of
Table~\ref{tab:coef}, respectively.

In Fig.~\ref{fig:comp}, we plot our numerical results together with the
contribution of the first two terms of the $\aZ$ expansion (dashed line). In
addition, we also plot the two-electron QED contribution, as evaluated by
Drake \cite{drake:CJP:88} (dotted line). It was obtained according
Eqs.~(2)-(9) of Ref.~\cite{drake:CJP:88}, keeping the contribution of first
order in $1/Z$ only. (We note that Eq.~(8) of Ref.~\cite{drake:CJP:88}
contains a misprint; its right-hand-side should be multiplied by $Z$.)
Expressions obtained in this way are exact to the leading order $\alpha^2
(\aZ)^3$. They also contain some higher-order contributions, due to all-order
results for the one-electron QED correction employed for the evaluation of
the $E_{L,1}$ term (Eq.~(2) of Ref.~\cite{drake:CJP:88}). We observe a good
agreement of our results with the previously known contributions and conclude
that Drake's values fall much closer to our all-order results than the pure
$\aZ$-expansion contribution.

For mixing states $(1s2p_{1/2})_1$ and $(1s2p_{3/2})_1$, Fig.~\ref{fig:comp}
presents a comparison for individual diagonal and off-diagonal matrix
elements. It should be mentioned that, generally speaking, comparison of
different methods should be performed for the physical energies obtained
after the diagonalization of the total matrix and not for the individual
matrix elements, since matrices with the same eigenvalues can differ by a
unitary transformation. We see from Fig.~\ref{fig:comp}, however, that our
results are in a good agreement with the $\aZ$-expansion contributions also
for the individual matrix elements.

An agreement found with the leading term of the $\aZ$ expansion offers us a
possibility to obtain the next-to-leading contribution, which is not known
analytically at present, and in this way to extend the results of our
calculations to lower values of $Z$. We thus isolate the higher-order
remainder $G^{\rm h.o.}_{\rm 2el}(\aZ)$ [see Eq.~(\ref{2elQED})] from our
numerical data and fit it to the form
\beq
G^{\rm h.o.}_{\rm 2el}(\aZ) =  a_{41}\ln \aZ+ a_{40}+(\aZ) (\ldots)\,.
\eeq
Fitted values for the coefficients $a_{41}$ and $a_{40}$ are presented in the
last two columns of Table~\ref{tab:coef}. It should be stressed that these
coefficients were obtained in the $jj$-coupling scheme with the wave
functions defined in case of mixing states by Eqs.~(\ref{u1}), (\ref{u2}).

There is a way to check the self-consistency of the numerical results for
individual matrix elements, which allows us to check each two-electron
QED contribution separately. We note that, in the LS coupling, the only
contribution to the off-diagonal matrix element to order $\alpha^2(\aZ)^3$ is
that of the anomalous magnetic moment correction $\Delta E_{\rm anom}$,
Eq.~(\ref{anom}). Therefore, for the two-photon exchange and screened
vacuum-polarization corrections, the off-diagonal matrix element in the LS
coupling is zero. In this case, the following identity is valid in the
jj-coupling scheme (to the order $\alpha^2(\aZ)^3$)
\beq    \label{id1}
\sqrt{2} [\Delta E_{(1s2p_{1/2})_1}-\Delta E_{(1s2p_{3/2})_1}] =
 -\Delta E_{\rm offdiag}^{jj}\,,
\eeq
where $\Delta E_i$ stand for the corresponding matrix elements. For the
screened self-energy correction, the off-diagonal matrix element in the LS
coupling ($\Delta E_{\rm offdiag}^{LS}$) is nonzero and
the corresponding identity reads
\beq    \label{id2}
\sqrt{2} [\Delta E_{(1s2p_{1/2})_1}-\Delta E_{(1s2p_{3/2})_1}] +
 \Delta E_{\rm offdiag}^{jj} = 3 \Delta E_{\rm offdiag}^{LS}\,.
\eeq
Fulfillment of these identities for individual two-electron QED contributions
is checked in Table~\ref{tab:identity}. For the screened self-energy and
vacuum-polarization correction, the fulfillment is obvious from the table.
For the two-photon exchange correction, the difference between the right- and
left-hand-side is very close to $3 (\aZ)^4$~eV in all cases listed and,
therefore, should be ascribed to higher-order contributions, for which the
identity is not valid anymore.

%%%%%%%%%%%%%%%%%%%%%%%%%%%%%%%%%%%%%%%%%%%%%%%%%%%%%%%%%%%%%%%%%%%%%%%%%
%
%
%
%%%%%%%%%%%%%%%%%%%%%%%%%%%%%%%%%%%%%%%%%%%%%%%%%%%%%%%%%%%%%%%%%%%%%%%%%
\section{Energies of $\bf n=1$ and $\bf n=2$ states of {He}-like ions}
\label{sec:compilation}

In this section we collect all contributions available to the ionization
energies of $n=1$ and $n=2$ states of He-like ions. Individual corrections
for selected ions are listed in Table~\ref{tab:He:contrib}. A description of
contributions presented there is given below.

{\em Dirac energy.} $\Delta E_{\rm Dirac}$ is the Dirac value for the
ionization energy of the valence electron including the finite-nuclear-size
effect. The energy levels were calculated employing the two-parameter Fermi
model for the nuclear-charge distribution. Parameters of the Fermi model were
expressed in terms of the root-mean-square (rms) radius (see, e.g.,
Ref.~\cite{shabaev:93:fns}), whose actual values were taken from
Refs.~\cite{fricke:95,vries:87,zumbro:84,zumbro:86}. For each value of $Z$,
the nuclear parameters for the isotope with the largest abundance (with the
longest life time) were chosen. An approximate formula from
Ref.~\cite{johnson:85} was employed for calculating rms radii for ions with
no experimental data available. In the table, we present also an estimation
of the uncertainty of the nuclear-size effect. In all cases except $Z=80$,
82, 83, 90, and 92, this uncertainty was evaluated by taking the one-percent
variation of the rms radius. For the above mentioned exceptions, the
rms-radii are supposed to be known more precisely. In our calculation we
employed the following values: 5.467(6)~Fm for $Z=80$, 5.504(25)~Fm for
$Z=82$, 5.533(20)~Fm for $Z=83$, 5.802(4)~Fm for $Z=90$, and 5.860(2)~Fm for
$Z=92$. The uncertainty of the nuclear-size effect in these cases was
evaluated by adding quadratically two errors, one obtained by varying the rms
radius within the error bars given and the other obtained by changing the
model of the nuclear-charge distribution (the Fermi and the
homogeneously-charged-sphere model were employed).

{\em Electron-electron interaction correction.} $\Delta E_{\rm int}$
incorporates corrections that can be derived from the Breit equation. It
consists of 3 parts,
\beq
\Delta E_{\rm int} = \Delta E_{\rm 1ph}+ \Delta E^{\rm Breit}_{\rm 2ph}+ \Delta
E^{\rm Breit}_{\ge{\rm 3ph}}\,,
\eeq
which correspond to the one, two, and three and more photon exchange,
respectively. In notations of Sec.~\ref{sec:derivation}, the one-photon
exchange correction is written as \cite{mittleman:72,shabaev:93}
\beq
\Delta E_{\rm 1ph} = \frac12 \sum_P(-1)^P\biggl[
I_{Pi_1Pi_2\,k_1k_2}(\Delta_1)+ I_{Pi_1Pi_2\,k_1k_2}(\Delta_2)\biggr]\,,
\eeq
where $\Delta_1 = \vare_{Pi_1}- \vare_{k_1}$ and $\Delta_2 = \vare_{Pi_2}-
\vare_{k_2}$. Its numerical evaluation was carried out employing the Fermi
model for the nuclear-charge distribution; accurate numerical results for
this correction can be found in Ref.~\cite{artemyev:00:pra}. $\Delta E^{\rm
Breit}_{\rm 2ph}$ represents the two-photon exchange correction within the
$\alpha^2(\aZ)^2$ approximation and is given by the first two terms in
Eq.~(\ref{2photQED}), with the coefficients $a_0$ and $a_2$ listed in
Table~\ref{tab:coef}. The contribution due to the exchange by three and more
photons was evaluated by summing terms of the $1/Z$ expansion, with the
corresponding coefficients taken from Refs.~\cite{sanders:69,aashamar:70} for
the nonrelativistic energy and from Ref.~\cite{drake:CJP:88} for the
Breit-Pauli correction.

{\em One-electron QED correction.} $\Delta E_{\rm 1el}^{\rm QED}$ is the sum
of the one-loop and two-loop one-electron QED corrections. The one-loop
self-energy correction for $1s$, $2s$, and $2p_{1/2}$ states and $Z\ge 26$
(including the nuclear-size effect) was tabulated in Ref.~\cite{beier:98:pra}
by using the method developed by Mohr and co-workers
\cite{mohr:74,mohr:92:a,mohr:93:prl}. For lower values of $Z$ and for the
$2p_{3/2}$ state, we used a combination of our own calculation and an
interpolation of the point-nucleus results from Ref.~\cite{mohr:92:b}. The
Uehling part of the one-loop vacuum-polarization correction was calculated in
this work for the Fermi nuclear model. The Wichmann-Kroll part of the
vacuum-polarization correction was tabulated for $Z\ge30$ in
Ref.~\cite{beier:97:jpb}. For lower values of $Z$, it was calculated in this
work by employing the asymptotic-expansion formulas for the Wichmann-Kroll
potential \cite{fainshtein:91}.

The two-loop one-electron QED correction is calculated to all orders in $\aZ$
only for the $1s$ state up to now, see Ref.~\cite{yerokhin:03:epjd} and
references therein. For excited states, one has to rely on the $\aZ$
expansion, which reads (see review \cite{mohr:00:rmp}, references therein,
and more recent studies \cite{pachucki:01:pra,pachucki:03:prl})
\beqn
\Delta E_{\rm 1el,2lo}^{\rm QED} &=& \frac{\alpha^2}{\pi^2}
 \frac{(\aZ)^4}{n^3}
    \left\{ B_{40}+ (\aZ)B_{50}+ (\aZ)^2 \left[L^3B_{63}+
    \right.\right. \nonumber \\ && \left.\left.
    L^2B_{62}
             +L\,B_{61}+G^{\rm h.o.}_{\rm 2lo}(\aZ)\right]  \right\}\,,
\eeqn
where $L = \ln [(\aZ)^{-2}]$, $G^{\rm h.o.}_{\rm 2lo}(\aZ) = B_{60}+
(\aZ)(\cdots)$ is the higher-order remainder, and the coefficients $B_{ij}$
are
\beqn
B_{40} &=& \left[
2\pi^2\ln2-\frac{49}{108}\pi^2-\frac{6131}{1296}-3\zeta(3)\right]\delta_{l0}
    \nonumber \\ &&
 +\left[ \frac12 \pi^2\ln2 -\frac1{12}\pi^2 -\frac{197}{144}-\frac34\zeta(3)\right]
    \frac1{\kappa(2l+1)}\,, \\
 B_{50} &=& -21.5561(31)\, \delta_{l0}\,, \\
 B_{63} &=& -\frac{8}{27}\, \delta_{l0}\,,
\eeqn
\beqn
 B_{62}(ns) &=& \frac{16}{9}\left( \frac{71}{60}-\ln(2n) +\frac1{4n^2}-\frac1n
             +\psi(n)+C \right)\,,  \\
 B_{62}(np) &=& \frac{4}{27}  \frac{n^2-1}{n^2}\,, \\
 B_{61}(1s) &=& 50.344005\,, \label{b611s} \\
 B_{61}(2s) &=& 42.447669\,, \label{b612s} \\
 B_{60}(1s) &=& -61.6(9)\,, \\
 B_{60}(2s) &=& -53.2(8)\,,
\eeqn
where $\zeta$ is the Riemann zeta function, $\psi$ is the logarithmic
derivative of the gamma function, and $C = 0.577261\ldots$ is the Euler
constant. Great care should be taken employing the $\aZ$ expansion for the
estimation of the total correction for middle- and high-$Z$ ions, due to a
very slow convergence of this expansion. In addition, it was found lately
\cite{yerokhin:04:jetpl} that the numerical all-order results do not agree
well with the analytical calculations to order $\alpha^2 (\aZ)^6$. A possible
reason for this disagreement \cite{pachucki:04:priv}
can be an incompleteness of the analytical
results (\ref{b611s}), (\ref{b612s}) for the $B_{61}$ coefficient.

In order to extrapolate the all-order numerical results of
Ref.~\cite{yerokhin:03:epjd} to the region $Z$=12-39 for the $1s$ state and
to estimate the two-loop correction for excited states, we separate the $1s$
higher-order remainder $G^{\rm h.o.}_{\rm 2lo}(\aZ)$ from the numerical data
of Ref.~\cite{yerokhin:03:epjd}. We observe that this function is smoothly
behaving and can be reasonably approximated by a polynomial. We thus employ a
linear (parabolic) fit to the function $G^{\rm h.o.}_{\rm 2lo}(\aZ)$ in order
to extrapolate the higher-order contribution to the region $Z$=12-39. For
$2s$ state, we employ the same values for the higher-order contribution and
ascribe the uncertainty of 50\% to them. For $p$ states, no analytical
calculations for the $B_{61}$ coefficient exist up to now. We thus separate
from the $1s$ numerical results of Ref.~\cite{yerokhin:03:epjd} the function
\beq
 \widetilde{G}^{\rm h.o.}_{\rm 2lo}(\aZ) = L\,B_{61}+
        G^{\rm h.o.}_{\rm 2lo}(\aZ)\,,
\eeq
divide it by a factor of 8, and take the result as the uncertainty for the
higher-order contribution for $p$ states.

{\em Two-electron QED correction.} $\Delta E_{\rm 2el}^{\rm QED}$ is
evaluated in Sec.~\ref{sec:twoelQED}; the data are taken from
Table~\ref{tab:twoelQED}.

{\em Higher-order QED correction.} $\Delta E_{\rm h.o.}^{\rm QED}$ represents
the contribution of QED effects of relative order $1/Z^2$ and higher. This
correction was evaluated by formulas presented in Ref.~\cite{drake:CJP:88}
suppressing terms that contribute to orders $1/Z^0$ and $1/Z$. Its
uncertainty was obtained by taking the relative deviation of the QED
contribution to order $1/Z$ calculated according to Ref.~\cite{drake:CJP:88}
from the results of its exact evaluation presented in this work. (The
corresponding comparison is presented in Fig.~\ref{fig:comp}.)

{\em Relativistic recoil correction.} $\Delta E_{\rm rec}$ consists of the
one-electron and the two-electron part. The one-electron relativistic recoil
correction was evaluated to all orders in $\aZ$ in a series of papers
\cite{artemyev:95:pra,artemyev:95:jpb,shabaev:98:recground}. In our
compilation, we employed the finite-nucleus results of
Ref.~\cite{shabaev:98:recground} for the $1s$ state, the point-nucleus
results of Ref.~\cite{artemyev:95:pra} for the $2s$ and $2p_{1/2}$ states,
and those of Ref.~\cite{artemyev:95:jpb} for the $2p_{3/2}$ state. The
two-electron recoil contribution is given by the sum of the mass-polarization
correction and the electron-electron interaction correction to the
one-electron nuclear recoil. The nonrelativistic part of the
mass-polarization correction was evaluated by summing the terms of the $1/Z$
expansion of the matrix element $\lbr \bm{p}_1\cdot\bm{p}_2\rbr$ taken from
Ref.~\cite{drake:CJP:88}. The known relativistic part of this correction of
order $(\aZ)^4m/M$ \cite{shabaev:94:rec} was also included. The
electron-electron interaction correction to the one-electron nuclear
recoil was taken into account in the nonrelativistic limit. It was estimated
as $(-m/M)\,\Delta E_{\rm 2el}$, where $\Delta E_{\rm 2el}$ is the total
two-electron correction.

In the last column of Table~\ref{tab:He:contrib} we present the total values
for the ionization energies, which are given by the sum of all corrections
mentioned so far. For lead, thorium, and uranium, the total values include
also the nuclear-polarization correction \cite{plunien:95,nefiodov:96}.
Analyzing the main sources of uncertainties listed in the table, we conclude
that in the low-$Z$ region the main error comes from the two-electron QED
corrections, namely from the two-photon exchange contribution. In the
high-$Z$ region, main sources of uncertainty are the one-electron two-loop
QED correction (mostly, the two-loop self-energy correction) and the
experimental values for the rms nuclear radii.

In Table~\ref{tab:total}, the total ionization energies of $n=1$ and $n=2$
states of He-like ions with $Z = 12-100$ are listed. We start our compilation
with $Z=12$ since this is the point where the new terms accounted for in our
calculation ($\sim \alpha^2(\aZ)^4$) become comparable with omitted
higher-order effects ($\sim \alpha^3 (\aZ)^2$).

In Fig.~\ref{fig:totcomp}, our results are compared with the theoretical
values obtained previously in calculations of different types
\cite{drake:CJP:88,plante:94,cheng:00}. Since our evaluation is the first one
complete to the order $\alpha^2 (\aZ)^4$, it is interesting to analyze the
difference between various calculations in units of $\alpha^2(\aZ)^4$. First
of all, we note a significant deviation of our values from the recent results
by Cheng and Chen \cite{cheng:00}, which arises from an incomplete treatment
of QED corrections employed in that work. The authors evaluate the QED
correction to all orders in $\aZ$ at the one-loop level, employing a
symmetric model potential in order to account for the electron-electron
interaction. This approximation works reasonably well in the high-$Z$ region,
but for ions with $22\le Z\le36$ (as presented in the paper), the accuracy of
this approximation turns out to be lower than that of Drake's approach based
on the exact $\aZ$ expansion \cite{drake:CJP:88}. We mention that a previous
investigation by these authors \cite{chen:93:pra} employed the QED correction
as evaluated by Drake. Its results agree well with those by Plante \etal$\,$
\cite{plante:94} and thus are in a better agreement with our numerical
values.

For the $^1S_0$ and $2\,^3P_{0,1}$ states, we observe also a distinct
deviation of our ionization energies from the results by Drake
\cite{drake:CJP:88}. A similar deviation was reported previously in the
literature \cite{johnson:92,chen:93:pra,plante:94}, where it was attributed
to corrections of order $\alpha^2(\aZ)^4$ to the electron-electron
interaction that were not accounted for by Drake's unified method but can be
(to a certain extent) included by methods based on the no-pair QED
Hamiltonian \cite{sucher:80}. Irregularities of the $Z$-dependence of the
plotted difference, which can be observed for $S$ states in the medium- and
high-$Z$ region, is explained by more recent values for the rms nuclear radii
employed in the present calculation.

As can be seen from Fig.~\ref{fig:totcomp}, the best agreement is found with
the calculation by Plante \etal$\,$ \cite{plante:94}. It is to be noted that
the results by Johnson and Sapirstein \cite{johnson:92} and by Chen \etal$\,$
\cite{chen:93:pra} obtained by different methods but on the same level of
sophistication are in a very good agreement with the ones by Plante and
co-workers. Whereas all these results are incomplete to order
$\alpha^2(\aZ)^4$, we conclude that the remaining contribution of this order
is rather small for all $n=2$ states, which explains a good agreement of
these results with the experimental data. Only for the $1\,^1S_1$ state, we
observe a significant new contribution of about 0.5$\,\alpha^2(\aZ)^4$. We
mention, however, that despite of a good agreement observed for the $n=2$
states, the results by Plante {\em et al.} are well outside of the estimated
error bars of the present theoretical values for most middle- and high-$Z$
ions.

In Table~\ref{tab:transen}, we list transition energies for which
experimental results are available. Comparison is made with the MBPT
calculation by Johnson and Sapirstein \cite{johnson:92}, with the CI
calculations by Chen \etal \cite{chen:93:pra}, and with the all-order
many-body treatment by Plante \etal \cite{plante:94}. These studies are,
according to our analysis, the most complete ones among the previous
calculations. We recall that in all these investigations QED corrections were
taken as evaluated by Drake \cite{drake:CJP:88}. The difference between them,
therefore, is related only to the part arising from the no-pair Hamiltonian,
often referred to as the ``structure" part.

We observe a generally good agreement of theoretical predictions with
experimental data. Despite of the significant amount of available
experimental information, the experimental uncertainty in the region of $Z$
under consideration is generally larger than the difference between the
calculations analyzed in Table~\ref{tab:transen}. Among few exceptions are
the recent high-precision measurements of the $2\,^3P_1-2\,^1S_0$ transition
energy in silicon ($Z=14$) \cite{redshaw:02} and the $2\,^3P_0-2\,^3P_1$
transition energy in magnesium ($Z=12$) \cite{myers:01}, whose accuracy is
much higher than that of the theoretical predictions. However, at these
relatively low values of $Z$, our treatment is basically equivalent to the
previous studies, and the difference between the calculations can not be
effectively probed in comparison with these measurements. When $Z$ increases,
deviation of our values from the results of the previous studies becomes more
prominent, but the experimental uncertainty is much lower for higher $Z$. A
compromise is found to be argon ($Z=18$), where the experimental
determination of the $2\,^3P_{0,2}-2\,^3S_1$ transition energies by Kukla
\etal \cite{kukla:95} demonstrated a 2$\sigma$ deviation from the previous
theoretical results. Our calculation brings the theoretical and experimental
results in agreement for the $2\,^3P_{0}-2\,^3S_1$ transition and reduces the
discrepancy for the $2\,^3P_{2}-2\,^3S_1$ transition to $0.5\,\sigma$.

An important feature of He-like ions is that they provide a possibility to
study the effects of parity non-conservation \cite{schaefer:89,karasiev:92}.
The $2\,^1S_0-2\,^3P_0$ transition in He-like Eu ion ($Z=63$) is presently
considered as the best candidate for future experiments \cite{labzowsky:01}.
The effect is enhanced by the fact that the $2\,^1S_0$ and $2\,^3P_0$ levels
cross each other in a vicinity of $Z=63$. Another crossing point of the
levels occurs around $Z=90$ but it seems to be less promising for the
experimental observation of the effect. In Table~\ref{tab:pncen} we list the
results of different theoretical evaluations for the $2\,^3P_0-2\,^1S_0$
transition energy in ions near the crossing points. A significant discrepancy
is observed between different theoretical evaluations, which is due to the
smallness of the energy difference for these ions. We mention a significant
deviation of our values from the recent results by Andreev \etal
\cite{andreev:03}. In that work, the authors performed an {\em ab initio}
calculation of the two-photon exchange correction, whose numerical values
agree well with those obtained in this paper. However, evaluating the total
transition energy, the authors used an estimation for the screened
self-energy correction (that was not calculated at that moment), which is the
main source of the disagreement observed.

Summarizing, in this investigation we performed {\it ab initio} QED
calculations of the screened self-energy correction and the two-photon
exchange correction for $n=1$ and $n=2$ states of He-like ions with $Z\ge
12$. This evaluation completes the rigorous treatment of all {\it
two-electron} QED corrections of order $\alpha^2$ to all orders in $\aZ$ and
significantly improves the theoretical accuracy for the energy values,
especially in the high-$Z$ region. Unlike all previous calculations, the
results obtained are complete through order $\alpha^2(\aZ)^4$; uncalculated
terms enter through three-photon-exchange QED effects ($\sim \alpha^3(\aZ)^2$
and higher) and through higher-order one-electron two-loop QED corrections
($\sim \alpha^2(\aZ)^7$ and higher).

%%%%%%%%%%%%%%%%%%%%%%%%%%%%%%%%%%%%%%%%%%%%%%%%%%%%%%%%%%%%%%%%%%%%%%%%%%%%%%
%
%%%%%%%%%%%%%%%%%%%%%%%%%%%%%%%%%%%%%%%%%%%%%%%%%%%%%%%%%%%%%%%%%%%%%%%%%%%%%%
\section*{Acknowledgements}

While finishing these investigations our friend and coauthor
Gerhard Soff deceased. His great penchant to the fundamental
aspects of the physics of strong fields and QED corrections
in heavy atoms made him a significant and inspiring driving
force in our collaboration, which can hardly be overestimated.
We shall miss him a lot.

Stimulating discussions with P. Mohr, J. Sapirstein, and T. St\"ohlker are
gratefully acknowledged. This work was supported in part by RFBR (Grant
No.~04-02-17574), by the Russian Ministry of Education (Grant
No.~E02-3.1-49), and by INTAS-GSI (grant No.~03-54-3604). The work of V.M.S.
was supported by the Alexander von Humboldt Stiftung. A.N.A. and
V.A.Y. acknowledge 
the support by the "Dynasty" foundation. 
The work of A.N.A. was also supported by
 INTAS YS grant No.  03-55-960 and by
Russian Ministry of Education and Administration of St. Petersburg (Grant
No.~PD02-1.2-79).
G.P. and G.S. acknowledge financial
support by the BMBF, DFG, and GSI. 

%%%%%%%%%%%%%%%%%%%%%%%%%%%%%%%%%%%
%
%
%           TABLES
%
%
%%%%%%%%%%%%%%%%%%%%%%%%%%%%%%%%%%%
\newpage
\begingroup
\begin{ruledtabular}
\begin{longtable*}{c........}
 \caption{The two-photon exchange correction for $n=1$ and $n=2$ states
of He-like ions, in eV. For mixing configurations, $(1s2p_{1/2})_1$ and
$(1s2p_{3/2})_1$ stand for the diagonal matrix elements of the operator $H$
[see Eqs.~(\ref{schr}), (\ref{schr0})], whereas "off-diag." labels the
off-diagonal matrix elements.
 \label{tab:2phot}}\\
\hline \hline
 $Z$  &\multicolumn{1}{c}{$(1s1s)_0$} & \multicolumn{1}{c}{$(1s2s)_0$} & \multicolumn{1}{c}{$(1s2s)_1$} & \multicolumn{1}{c}{$(1s2p_{1/2})_0$}&\multicolumn{1}{c}{$(1s2p_{1/2})_1$}&\multicolumn{1}{c}{$(1s2p_{3/2})_1$}&\multicolumn{1}{c}{$(1s2p_{3/2})_2$}&  \multicolumn{1}{c}{off-diag.} \\
\hline
 12 & -4.4x186     & -3.1x741     & -1.2x991     & -2.0x506     & -2.7x789     & -3.5x332     & -1.9x964     & -1.0x711        \\
 14 & -4.4x645     & -3.1x952     & -1.3x024     & -2.0x741     & -2.7x899     & -3.5x413     & -2.0x000     & -1.0x686        \\
    &     x        & -3.1x9541^b  & -1.3x0240^b  &     x        &     x        &     x        &     x        &     x           \\
 16 & -4.5x173     & -3.2x196     & -1.3x062     & -2.1x015     & -2.8x027     & -3.5x507     & -2.0x041     & -1.0x658        \\
 18 & -4.5x770     & -3.2x473     & -1.3x106     & -2.1x328     & -2.8x173     & -3.5x613     & -2.0x088     & -1.0x626        \\
    &     x        & -3.2x4753^b  & -1.3x1057^b  &     x        & -2.8x168^e   & -3.5x603^e   &     x        & -1.0x618^e      \\
 20 & -4.6x435     & -3.2x784     & -1.3x154     & -2.1x682     & -2.8x337     & -3.5x733     & -2.0x141     & -1.0x589        \\
    & -4.6x447^a   &     x        &     x        &     x        &     x        &     x        &     x        &     x           \\
 28 & -4.9x784     & -3.4x378     & -1.3x405     & -2.3x532     & -2.9x182     & -3.6x340     & -2.0x405     & -1.0x406        \\
 30 & -5.0x795     & -3.4x868     & -1.3x483     & -2.4x111     & -2.9x443     & -3.6x525     & -2.0x484     & -1.0x350        \\
    & -5.0x812^a   & -3.4x8716^b  & -1.3x4827^b  & -2.4x1112^d  & -2.9x439^e   & -3.6x506^e   & -2.0x4834^d  & -1.0x350^e      \\
    &     x        & -3.4x73^c    & -1.3x48^c    &     x        &     x        &     x        &     x        &     x           \\
    &     x        &     x        & -1.3x4833^d  &     x        &     x        &     x        &     x        &     x           \\
 32 & -5.1x877     & -3.5x396     & -1.3x566     & -2.4x741     & -2.9x725     & -3.6x724     & -2.0x568     & -1.0x291        \\
 40 & -5.6x924     & -3.7x919     & -1.3x961     & -2.7x817     & -3.1x072     & -3.7x658     & -2.0x956     & -1.0x015        \\
    & -5.6x945^a   &     x        & -1.3x9621^d  & -2.7x8172^d  & -3.1x082^e   & -3.7x641^e   & -2.0x9545^d  & -1.0x008^e      \\
 47 & -6.2x332     & -4.0x719     & -1.4x395     & -3.1x351     & -3.2x575     & -3.8x668     & -2.1x358     & -0.9x724        \\
 50 & -6.4x951     & -4.2x110     & -1.4x609     & -3.3x148     & -3.3x323     & -3.9x159     & -2.1x548     & -0.9x586        \\
    & -6.4x975^a   &     x        & -1.4x6120^d  & -3.3x1489^d  & -3.3x33^e    & -3.9x15^e    & -2.1x5465^d  & -0.9x55^e       \\
 54 & -6.8x742     & -4.4x162     & -1.4x923     & -3.5x848     & -3.4x429     & -3.9x871     & -2.1x816     & -0.9x387        \\
 60 & -7.5x114     & -4.7x714     & -1.5x459     & -4.0x642     & -3.6x348     & -4.1x066     & -2.2x251     & -0.9x064        \\
    & -7.5x142^a   & -4.7x7215^b  & -1.5x4587^b  & -4.0x68^c    & -3.6x35^e    & -4.1x05^e    & -2.2x2510^d  & -0.8x93^e       \\
    &     x        & -4.7x81^c    & -1.5x42^c    & -4.0x6446^d  &     x        &     x        &     x        &     x           \\
    &     x        &     x        & -1.5x4558^d  &     x        &     x        &     x        &     x        &     x           \\
 66 & -8.2x393     & -5.1x924     & -1.6x082     & -4.6x505     & -3.8x632     & -4.2x426     & -2.2x724     & -0.8x708        \\
    &     x        & -5.1x94^c    & -1.6x05^c    & -4.6x70^c    &     x        &     x        &     x        &     x           \\
 70 & -8.7x812     & -5.5x159     & -1.6x552     & -5.1x131     & -4.0x394     & -4.3x430     & -2.3x060     & -0.8x453        \\
    & -8.7x847^a   & -5.5x15^c    & -1.6x48^c    & -5.1x17^c    & -4.0x38^e    & -4.3x39^e    & -2.3x0573^d  & -0.8x01^e       \\
    &     x        &     x        & -1.6x5478^d  & -5.1x1403^d  &     x        &     x        &     x        &     x           \\
 74 & -9.3x739     & -5.8x794     & -1.7x073     & -5.6x441     & -4.2x381     & -4.4x517     & -2.3x412     & -0.8x184        \\
 79 &-10.1x957     & -6.3x996     & -1.7x803     & -6.4x220     & -4.5x238     & -4.5x999     & -2.3x877     & -0.7x826        \\
 80 &-10.3x719     & -6.5x135     & -1.7x961     & -6.5x950     & -4.5x866     & -4.6x312     & -2.3x974     & -0.7x752        \\
    &-10.3x75^a    & -6.5x04^c    & -1.7x89^c    & -6.5x98^c    & -4.5x85^e    & -4.6x28^e    & -2.3x9806^d  & -0.7x71^e       \\
    &     x        &     x        & -1.7x9562^d  & -6.5x9593^d  &     x        &     x        &     x        &     x           \\
 82 &-10.7x375     & -6.7x524     & -1.8x289     & -6.9x607     & -4.7x185     & -4.6x957     & -2.4x170     & -0.7x601        \\
 83 &-10.9x271     & -6.8x776     & -1.8x460     & -7.1x540     & -4.7x877     & -4.7x288     & -2.4x270     & -0.7x524        \\
 90 &-12.3x979     & -7.8x792     & -1.9x790     & -8.7x331     & -5.3x458     & -4.9x780     & -2.5x005     & -0.6x957        \\
    &-12.4x03^a    &     x        &     x        &     x        &     x        &     x        &     x        &     x           \\
 92 &-12.8x714     & -8.2x122     & -2.0x221     & -9.2x701     & -5.5x329     & -5.0x550     & -2.5x228     & -0.6x787        \\
    &     x        & -8.2x1306^b  & -2.0x2199^b  & -9.2x74^c    & -5.5x31^e    & -5.0x53^e    & -2.5x2228^d  & -0.6x83^e       \\
    &     x        & -8.1x84^c    & -2.0x18^c    & -9.2x7598^d  &     x        &     x        &     x        &     x           \\
    &     x        &     x        & -2.0x2034^d  &     x        &     x        &     x        &     x        &     x           \\
100 &-15.0x772     & -9.8x239     & -2.2x223     &-11.9x330     & -6.4x484     & -5.3x900     & -2.6x191     & -0.6x058        \\
    &-15.0x805^a   &     x        &     x        &     x        &     x        &     x        &     x        &     x           \\
\hline \hline
\end{longtable*}
\end{ruledtabular}
 $^a$ Blundell \etal \cite{blundell:93:b},
 $^b$ \r{A}sen \etal \cite{asen:02},
 $^c$ Andreev \etal \cite{andreev:01,andreev:03},
 $^d$ Mohr and Sapirstein \cite{mohr:00:pra},
 $^e$ Andreev \etal \cite{andreev:04}.
\endgroup

%%%%%%%%%%%%%%%%%%%%%%%%%%%%%%%%%%%%%%%%%%%%%%%%%%%%%%%%%%%%%%%%%%%%%%%%
%
%
%
%%%%%%%%%%%%%%%%%%%%%%%%%%%%%%%%%%%%%%%%%%%%%%%%%%%%%%%%%%%%%%%%%%%%%%%%
\begingroup
\squeezetable
\begin{table} \caption{Screened self-energy correction for $n=1$ and $n=2$ states
of He-like ions, in units of $F(\aZ)$. In case of mixing configurations,
contributions to the matrix elements $H_{ik}$ are given; labeling is as in
Table~\ref{tab:2phot}.
 \label{tab:sescr}}
\begin{ruledtabular}
\begin{tabular}{r........}
$Z$ &  \multicolumn{1}{c}{$(1s1s)_0$}
                     & \multicolumn{1}{c}{$(1s2s)_0$}
                                      & \multicolumn{1}{c}{$(1s2s)_1$}
                                                       &  \multicolumn{1}{c}{$(1s2p_{1/2})_0$}
                                                                        & \multicolumn{1}{c}{$(1s2p_{1/2})_1$}
                                                                                          & \multicolumn{1}{c}{$(1s2p_{3/2})_1$}
                                                                                                            & \multicolumn{1}{c}{$(1s2p_{3/2})_2$}
                                                                                                                              &  \multicolumn{1}{c}{off-diag.}\\
\hline
 12 &  -2.21x39(8)   &   -0.48x41(5)  &   -0.30x31(5)  &  -0.09x17(6)   &   -0.06x91(6)   &   -0.05x56(7)   &   -0.13x50(7)   &   0.05x33(2)     \\
 14 &  -2.05x43(6)   &   -0.45x19(4)  &   -0.28x21(4)  &  -0.08x45(5)   &   -0.06x46(5)   &   -0.05x37(6)   &   -0.12x66(6)   &   0.04x90(1)     \\
 16 &  -1.92x17(3)   &   -0.42x48(3)  &   -0.26x46(3)  &  -0.07x83(4)   &   -0.06x05(4)   &   -0.05x17(4)   &   -0.11x97(4)   &   0.04x559(5)    \\
 18 &  -1.80x97(3)   &   -0.40x21(3)  &   -0.24x96(3)  &  -0.07x33(2)   &   -0.05x71(2)   &   -0.05x01(2)   &   -0.11x37(2)   &   0.04x266(3)    \\
 20 &  -1.71x37(3)   &   -0.38x28(3)  &   -0.23x68(3)  &  -0.06x93(1)   &   -0.05x44(1)   &   -0.04x88(2)   &   -0.10x86(2)   &   0.04x013(3)    \\
 30 &  -1.38x88(2)   &   -0.31x94(2)  &   -0.19x30(2)  &  -0.05x81(1)   &   -0.04x70(1)   &   -0.04x52(2)   &   -0.09x13(2)   &   0.03x146(2)    \\
 40 &  -1.21x12(1)   &   -0.28x79(1)  &   -0.16x85(1)  &  -0.05x588(7)  &   -0.04x542(7)  &   -0.04x42(1)   &   -0.08x17(1)   &   0.02x639(2)    \\
 50 &  -1.11x34(1)   &   -0.27x46(1)  &   -0.15x47(1)  &  -0.05x963(8)  &   -0.04x784(8)  &   -0.04x49(1)   &   -0.07x61(1)   &   0.02x312(2)    \\
 60 &  -1.06x79(1)   &   -0.27x44(1)  &   -0.14x78(1)  &  -0.06x871(6)  &   -0.05x371(6)  &   -0.04x65(1)   &   -0.07x29(1)   &   0.02x087(1)    \\
 70 &  -1.06x281(5)  &   -0.28x559(5) &   -0.14x670(5) &  -0.08x394(5)  &   -0.06x349(5)  &   -0.04x896(7)  &   -0.07x136(7)  &   0.01x9257(8)   \\
 80 &  -1.09x510(3)  &   -0.30x916(3) &   -0.15x096(3) &  -0.10x779(2)  &   -0.07x864(2)  &   -0.05x197(7)  &   -0.07x091(7)  &   0.01x8047(6)   \\
 83 &  -1.11x237(2)  &   -0.31x903(2) &   -0.15x336(2) &  -0.11x728(2)  &   -0.08x463(2)  &   -0.05x294(7)  &   -0.07x094(7)  &   0.01x7741(5)   \\
 90 &  -1.16x760(2)  &   -0.34x804(2) &   -0.16x122(2) &  -0.14x526(1)  &   -0.10x222(1)  &   -0.05x530(7)  &   -0.07x130(7)  &   0.01x7104(3)   \\
 92 &  -1.18x776(2)  &   -0.35x814(2) &   -0.16x413(2) &  -0.15x515(1)  &   -0.10x841(1)  &   -0.05x600(7)  &   -0.07x148(7)  &   0.01x6939(3)   \\
100 &  -1.29x293(2)  &   -0.40x917(2) &   -0.17x942(2) &  -0.20x688(3)  &   -0.14x073(3)  &   -0.05x881(7)  &   -0.07x250(7)  &   0.01x6343(3)   \\
\end{tabular}
\end{ruledtabular}
\end{table}
\endgroup

%%%%%%%%%%%%%%%%%%%%%%%%%%%%%%%%%%%%%%%%%%%%%%%%%%%%%%%%%%%%%%%%%%%%%%%%
%
%
%
%%%%%%%%%%%%%%%%%%%%%%%%%%%%%%%%%%%%%%%%%%%%%%%%%%%%%%%%%%%%%%%%%%%%%%%%
\begingroup
\begin{table}
\caption{Coefficients of the $\aZ$ expansion of the
second-order two-electron contribution
to the energy levels of He-like ions. In case of mixing
configurations, contributions to the matrix elements $H_{ik}$ are given;
labeling is as in Table~\ref{tab:2phot}.
 \label{tab:coef}}
\begin{ruledtabular}
\begin{tabular}{l......}
   & \multicolumn{1}{c}{$(\aZ)^0$} &
                \multicolumn{1}{c}{$(\aZ)^2$} &
                     \multicolumn{1}{c}{$(\aZ)^3\,\ln \aZ$} &
                                \multicolumn{1}{c}{$(\aZ)^3$}&
                                          \multicolumn{1}{c}{$(\aZ)^4\,\ln \aZ$} &
                                                         \multicolumn{1}{c}{$(\aZ)^4$} \\
\hline
$(1s1s)_0$       &  -0.15x7662  & -0.6x302 & 1.3x191 & 1.6x588 &    0.7x5(15)  &  -2.4x1(40) \\
$(1s2s)_0$       &  -0.11x4509  & -0.2x807 & 0.2x755 & 0.3x255 &    0.1x1(2)   &  -0.8x1(5)  \\
$(1s2s)_1$       &  -0.04x7409  & -0.0x428 & 0.1x795 & 0.1x911 &    0.0x56(11) &  -0.4x0(3)  \\
$(1s2p_{1/2})_0$ &  -0.07x2999  & -0.3x035 & 0.0x730 & 0.1x063 &    0          &  -0.6x4(2)  \\
$(1s2p_{1/2})_1$ &  -0.10x1008  & -0.1x444 & 0.0x465 & 0.0x578 &    0          &  -0.2x2(1)  \\
$(1s2p_{3/2})_1$ &  -0.12x9018  & -0.1x075 & 0.0x201 & 0.0x058 &    0          &  -0.1x0     \\
$(1s2p_{3/2})_2$ &  -0.07x2999  & -0.0x473 & 0.0x730 & 0.0x595 &    0.0x1(1)   &  -0.1x4(2)  \\
off-diag.        &  -0.03x9611  &  0.0x319 &-0.0x374 &-0.0x432 &   -0.0x1(1)   &   0.0x8(4)  \\
\end{tabular}
\end{ruledtabular}
\end{table}
\endgroup

%%%%%%%%%%%%%%%%%%%%%%%%%%%%%%%%%%%%%%%%%%%%%%%%%%%%%%%%%%%%%%%%%%%%
%
%
%
%%%%%%%%%%%%%%%%%%%%%%%%%%%%%%%%%%%%%%%%%%%%%%%%%%%%%%%%%%%%%%%%%%%%
\begingroup
\squeezetable
\begin{ruledtabular}
\begin{longtable*}{rl....||rl....}
\caption{Two-electron QED correction for $n=1$ and $n = 2$ states of
He-like ions, in eV.
    \label{tab:twoelQED}
    }\\
\colrule\hline
$Z$ & State & \multicolumn{1}{c}{Scr.SE} & \multicolumn{1}{c}{Scr.VP} & \multicolumn{1}{c}{2-ph.exch.} & \multicolumn{1}{c}{Total} &
$Z$ & State & \multicolumn{1}{c}{Scr.SE} & \multicolumn{1}{c}{Scr.VP} & \multicolumn{1}{c}{2-ph.exch.} & \multicolumn{1}{c}{Total}
\\    \hline
 12 & $(1s)^2$         &   -0.04x05    &    0.0x021     &    0.00x31(1)  &   -0.03x53(1)  &  60 & $(1s)^2$         &   -2.43x92(2) &    0.38x00(1)  &    0.06x62(1)  &   -1.99x30(2) \\
    & $(1s2s)_0$       &   -0.00x88    &    0.0x004     &    0.00x04(1)  &   -0.00x80(1)  &     & $(1s2s)_0$       &   -0.62x67(2) &    0.09x23     &   -0.19x14(1)  &   -0.72x58(2) \\
    & $(1s2s)_1$       &   -0.00x55    &    0.0x003     &   -0.00x01(1)  &   -0.00x53(1)  &     & $(1s2s)_1$       &   -0.33x77(2) &    0.04x84     &   -0.03x27(1)  &   -0.32x19(2) \\
    & $(1s2p_{1/2})_0$ &   -0.00x17    &    0.0x001     &   -0.00x08(1)  &   -0.00x24(1)  &     & $(1s2p_{1/2})_0$ &   -0.15x69(1) &    0.03x11     &   -0.49x45(1)  &   -0.62x03(1) \\
    & $(1s2p_{1/2})_1$ &   -0.00x13    &    0.0x001     &   -0.00x02(1)  &   -0.00x14(1)  &     & $(1s2p_{1/2})_1$ &   -0.12x27(1) &    0.01x90     &   -0.13x30(1)  &   -0.23x67(1) \\
    & $(1s2p_{3/2})_1$ &   -0.00x10    &    0.0x000     &    0.00x00(1)  &   -0.00x10(1)  &     & $(1s2p_{3/2})_1$ &   -0.10x63(2) &    0.00x46     &   -0.03x49(1)  &   -0.13x66(2) \\
    & $(1s2p_{3/2})_2$ &   -0.00x25    &    0.0x001     &   -0.00x01(1)  &   -0.00x24(1)  &     & $(1s2p_{3/2})_2$ &   -0.16x66(2) &    0.01x59     &    0.00x81(1)  &   -0.14x26(2) \\
    &   off-diag.      &    0.00x10    &   -0.0x001     &    0.00x01(1)  &    0.00x10(1)  &     &   off-diag.      &    0.04x77    &   -0.00x92     &    0.00x53(2)  &    0.04x37(2) \\
 \hline
 14 & $(1s)^2$         &   -0.05x96    &    0.0x034     &    0.00x46(1)  &   -0.05x16(1)  &  70 & $(1s)^2$         &   -3.85x48(1) &    0.71x30(2)  &   -0.01x64(1)  &   -3.15x81(2) \\
    & $(1s2s)_0$       &   -0.01x31    &    0.0x007     &    0.00x05(1)  &   -0.01x19(1)  &     & $(1s2s)_0$       &   -1.03x58(1) &    0.18x19     &   -0.40x71(1)  &   -1.26x10(2) \\
    & $(1s2s)_1$       &   -0.00x82    &    0.0x005     &   -0.00x02(1)  &   -0.00x79(1)  &     & $(1s2s)_1$       &   -0.53x21(1) &    0.08x92     &   -0.06x15(1)  &   -0.50x43(2) \\
    & $(1s2p_{1/2})_0$ &   -0.00x25    &    0.0x002     &   -0.00x15(1)  &   -0.00x37(1)  &     & $(1s2p_{1/2})_0$ &   -0.30x44(1) &    0.06x67     &   -0.97x17(1)  &   -1.20x94(2) \\
    & $(1s2p_{1/2})_1$ &   -0.00x19    &    0.0x001     &   -0.00x04(1)  &   -0.00x21(1)  &     & $(1s2p_{1/2})_1$ &   -0.23x03(1) &    0.04x09     &   -0.26x56(1)  &   -0.45x50(2) \\
    & $(1s2p_{3/2})_1$ &   -0.00x16    &    0.0x001     &    0.00x00(1)  &   -0.00x15(1)  &     & $(1s2p_{3/2})_1$ &   -0.17x76(2) &    0.00x75     &   -0.06x88(1)  &   -0.23x88(2) \\
    & $(1s2p_{3/2})_2$ &   -0.00x37    &    0.0x002     &   -0.00x01(1)  &   -0.00x36(1)  &     & $(1s2p_{3/2})_2$ &   -0.25x88(2) &    0.02x66     &    0.01x64(1)  &   -0.21x58(2) \\
    &   off-diag.      &    0.00x14    &   -0.0x001     &    0.00x02(1)  &    0.00x15(1)  &     &   off-diag.      &    0.06x98    &   -0.01x58     &    0.00x63(2)  &    0.06x03(2) \\
 \hline
 16 & $(1s)^2$         &   -0.08x32    &    0.0x051     &    0.00x66(1)  &   -0.07x15(1)  &  80 & $(1s)^2$         &   -5.92x89(1) &    1.29x80(2)  &   -0.23x74(1)  &   -4.86x82(3) \\
    & $(1s2s)_0$       &   -0.01x84    &    0.0x011     &    0.00x05(1)  &   -0.01x68(1)  &     & $(1s2s)_0$       &   -1.67x38(1) &    0.35x20(1)  &   -0.79x46(1)  &   -2.11x64(2) \\
    & $(1s2s)_1$       &   -0.01x15    &    0.0x007     &   -0.00x03(1)  &   -0.01x10(1)  &     & $(1s2s)_1$       &   -0.81x73(1) &    0.16x15(1)  &   -0.10x93(1)  &   -0.76x52(2) \\
    & $(1s2p_{1/2})_0$ &   -0.00x34    &    0.0x003     &   -0.00x25(1)  &   -0.00x56(1)  &     & $(1s2p_{1/2})_0$ &   -0.58x36(1) &    0.14x29     &   -1.79x38(1)  &   -2.23x45(1) \\
    & $(1s2p_{1/2})_1$ &   -0.00x26    &    0.0x002     &   -0.00x06(1)  &   -0.00x30(1)  &     & $(1s2p_{1/2})_1$ &   -0.42x58(1) &    0.08x79     &   -0.49x88(1)  &   -0.83x67(1) \\
    & $(1s2p_{3/2})_1$ &   -0.00x22    &    0.0x001     &    0.00x00(1)  &   -0.00x22(1)  &     & $(1s2p_{3/2})_1$ &   -0.28x14(3) &    0.01x20     &   -0.12x32(1)  &   -0.39x26(3) \\
    & $(1s2p_{3/2})_2$ &   -0.00x52    &    0.0x003     &   -0.00x02(1)  &   -0.00x51(1)  &     & $(1s2p_{3/2})_2$ &   -0.38x39(3) &    0.04x28     &    0.02x79(1)  &   -0.31x32(3) \\
    &   off-diag.      &    0.00x20    &   -0.0x002     &    0.00x03(1)  &    0.00x21(1)  &     &   off-diag.      &    0.09x77    &   -0.02x60     &    0.00x72(2)  &    0.07x89(2) \\
 \hline
 18 & $(1s)^2$         &   -0.11x16    &    0.0x072     &    0.00x91(1)  &   -0.09x53(1)  &  83 & $(1s)^2$         &   -6.72x56(1) &    1.55x00(7)  &   -0.34x60(1)  &   -5.52x16(7) \\
    & $(1s2s)_0$       &   -0.02x48    &    0.0x015     &    0.00x04(1)  &   -0.02x28(1)  &     & $(1s2s)_0$       &   -1.92x89(1) &    0.42x86(2)  &   -0.95x99(1)  &   -2.46x02(2) \\
    & $(1s2s)_1$       &   -0.01x54    &    0.0x010     &   -0.00x04(1)  &   -0.01x48(1)  &     & $(1s2s)_1$       &   -0.92x73(1) &    0.19x27(2)  &   -0.12x89(1)  &   -0.86x35(2) \\
    & $(1s2p_{1/2})_0$ &   -0.00x45    &    0.0x004     &   -0.00x39(1)  &   -0.00x80(1)  &     & $(1s2p_{1/2})_0$ &   -0.70x91(1) &    0.17x99(1)  &   -2.13x77(1)  &   -2.66x69(1) \\
    & $(1s2p_{1/2})_1$ &   -0.00x35    &    0.0x003     &   -0.00x10(1)  &   -0.00x42(1)  &     & $(1s2p_{1/2})_1$ &   -0.51x17(1) &    0.11x09(1)  &   -0.59x77(1)  &   -0.99x85(1) \\
    & $(1s2p_{3/2})_1$ &   -0.00x31    &    0.0x001     &   -0.00x01(1)  &   -0.00x31(1)  &     & $(1s2p_{3/2})_1$ &   -0.32x01(4) &    0.01x36     &   -0.14x46(1)  &   -0.45x11(4) \\
    & $(1s2p_{3/2})_2$ &   -0.00x70    &    0.0x004     &   -0.00x02(1)  &   -0.00x68(1)  &     & $(1s2p_{3/2})_2$ &   -0.42x89(4) &    0.04x89     &    0.03x18(1)  &   -0.34x82(4) \\
    &   off-diag.      &    0.00x26    &   -0.0x002     &    0.00x04(1)  &    0.00x28(1)  &     &   off-diag.      &    0.10x73    &   -0.02x99     &    0.00x74(4)  &    0.08x48(4) \\
 \hline
 20 & $(1s)^2$         &   -0.14x50    &    0.0x099     &    0.01x19(1)  &   -0.12x31(1)  &  90 & $(1s)^2$         &   -9.00x06(1) &    2.33x8(1)   &   -0.71x09(1)  &   -7.37x3(1)  \\
    & $(1s2s)_0$       &   -0.03x24    &    0.0x021     &    0.00x03(1)  &   -0.03x00(1)  &     & $(1s2s)_0$       &   -2.68x29(1) &    0.68x10(2)  &   -1.46x89(1)  &   -3.47x08(3) \\
    & $(1s2s)_1$       &   -0.02x00    &    0.0x014     &   -0.00x06(1)  &   -0.01x92(1)  &     & $(1s2s)_1$       &   -1.24x28(1) &    0.29x21(2)  &   -0.18x69(1)  &   -1.13x76(2) \\
    & $(1s2p_{1/2})_0$ &   -0.00x59    &    0.0x006     &   -0.00x59(1)  &   -0.01x11(1)  &     & $(1s2p_{1/2})_0$ &   -1.11x97    &    0.31x12(2)  &   -3.18x42(1)  &   -3.99x28(2) \\
    & $(1s2p_{1/2})_1$ &   -0.00x46    &    0.0x004     &   -0.00x15(1)  &   -0.00x57(1)  &     & $(1s2p_{1/2})_1$ &   -0.78x79    &    0.19x29(1)  &   -0.90x24(1)  &   -1.49x74(1) \\
    & $(1s2p_{3/2})_1$ &   -0.00x41    &    0.0x001     &   -0.00x02(1)  &   -0.00x42(1)  &     & $(1s2p_{3/2})_1$ &   -0.42x63(5) &    0.01x76     &   -0.20x51(1)  &   -0.61x38(5) \\
    & $(1s2p_{3/2})_2$ &   -0.00x92    &    0.0x006     &   -0.00x03(1)  &   -0.00x89(1)  &     & $(1s2p_{3/2})_2$ &   -0.54x96(5) &    0.06x63     &    0.04x13(1)  &   -0.44x20(5) \\
    &   off-diag.      &    0.00x34    &   -0.0x003     &    0.00x05(1)  &    0.00x36(1)  &     &   off-diag.      &    0.13x18    &   -0.04x10     &    0.00x82(4)  &    0.09x91(4) \\
 \hline
 30 & $(1s)^2$         &   -0.39x65    &    0.0x348     &    0.03x25(1)  &   -0.32x92(1)  &  92 & $(1s)^2$         &   -9.78x00(1) &    2.63x0(2)   &   -0.85x20(1)  &   -8.00x2(2) \\
    & $(1s2s)_0$       &   -0.09x12    &    0.0x076     &   -0.00x48(1)  &   -0.08x84(1)  &     & $(1s2s)_0$       &   -2.94x89(1) &    0.77x70(4)  &   -1.65x40(1)  &   -3.82x59(4) \\
    & $(1s2s)_1$       &   -0.05x51    &    0.0x048     &   -0.00x24(1)  &   -0.05x27(1)  &     & $(1s2s)_1$       &   -1.35x14(1) &    0.32x87(2)  &   -0.20x74(1)  &   -1.23x01(3) \\
    & $(1s2p_{1/2})_0$ &   -0.01x66    &    0.0x022     &   -0.02x89(1)  &   -0.04x33(1)  &     & $(1s2p_{1/2})_0$ &   -1.27x75    &    0.36x47(2)  &   -3.56x12(1)  &   -4.47x40(3) \\
    & $(1s2p_{1/2})_1$ &   -0.01x34    &    0.0x013     &   -0.00x74(1)  &   -0.01x95(1)  &     & $(1s2p_{1/2})_1$ &   -0.89x27    &    0.22x62(2)  &   -1.01x33(1)  &   -1.67x98(2) \\
    & $(1s2p_{3/2})_1$ &   -0.01x29    &    0.0x005     &   -0.00x16(1)  &   -0.01x40(1)  &     & $(1s2p_{3/2})_1$ &   -0.46x11(5) &    0.01x88     &   -0.22x54(1)  &   -0.66x77(5) \\
    & $(1s2p_{3/2})_2$ &   -0.02x61    &    0.0x019     &   -0.00x03(1)  &   -0.02x45(1)  &     & $(1s2p_{3/2})_2$ &   -0.58x86(5) &    0.07x21     &    0.04x40(1)  &   -0.47x25(5) \\
    &   off-diag.      &    0.00x90    &   -0.0x010     &    0.00x13(1)  &    0.00x93(1)  &     &   off-diag.      &    0.13x95    &   -0.04x48     &    0.00x84(4)  &    0.10x31(4) \\
 \hline
 40 & $(1s)^2$         &   -0.81x97    &    0.0x887     &    0.05x89(1)  &   -0.67x21(1)  & 100 & $(1s)^2$         &  -13.67x16(1) &    4.24x8(4)   &   -1.65x51(1)  &  -11.07x9(4)  \\
    & $(1s2s)_0$       &   -0.19x48    &    0.0x199     &   -0.02x52(1)  &   -0.20x02(1)  &     & $(1s2s)_0$       &   -4.32x66(1) &    1.34x04(8)  &   -2.64x09(1)  &   -5.62x71(8) \\
    & $(1s2s)_1$       &   -0.11x41    &    0.0x118     &   -0.00x68(1)  &   -0.10x91(1)  &     & $(1s2s)_1$       &   -1.89x72(1) &    0.53x66(5)  &   -0.31x24(1)  &   -1.67x30(5) \\
    & $(1s2p_{1/2})_0$ &   -0.03x78    &    0.0x060     &   -0.09x16(1)  &   -0.12x34(1)  &     & $(1s2p_{1/2})_0$ &   -2.18x76(3) &    0.70x67(5)  &   -5.54x84(1)  &   -7.02x93(6) \\
    & $(1s2p_{1/2})_1$ &   -0.03x07    &    0.0x036     &   -0.02x39(1)  &   -0.05x10(1)  &     & $(1s2p_{1/2})_1$ &   -1.48x81(3) &    0.44x08(5)  &   -1.60x74(1)  &   -2.65x47(6) \\
    & $(1s2p_{3/2})_1$ &   -0.02x99    &    0.0x012     &   -0.00x58(1)  &   -0.03x45(1)  &     & $(1s2p_{3/2})_1$ &   -0.62x19(7) &    0.02x34     &   -0.32x11(1)  &   -0.91x95(7) \\
    & $(1s2p_{3/2})_2$ &   -0.05x53    &    0.0x044     &    0.00x05(1)  &   -0.05x03(1)  &     & $(1s2p_{3/2})_2$ &   -0.76x66(7) &    0.10x09(1)  &    0.05x29(1)  &   -0.61x28(7) \\
    &   off-diag.      &    0.01x79    &   -0.0x025     &    0.00x25(1)  &    0.01x79(1)  &     &   off-diag.      &    0.17x28    &   -0.06x30     &    0.01x04(4)  &    0.12x02(4) \\
 \hline
 50 & $(1s)^2$         &   -1.47x17(1) &    0.1x920     &    0.07x81(1)  &   -1.20x16(1) \\
    & $(1s2s)_0$       &   -0.36x30(1) &    0.0x446     &   -0.07x83(1)  &   -0.39x66(1) \\
    & $(1s2s)_1$       &   -0.20x44(1) &    0.0x250     &   -0.01x59(1)  &   -0.19x53(1) \\
    & $(1s2p_{1/2})_0$ &   -0.07x88(1) &    0.0x141     &   -0.22x89(1)  &   -0.29x36(1) \\
    & $(1s2p_{1/2})_1$ &   -0.06x32(1) &    0.0x086     &   -0.06x06(1)  &   -0.11x52(1) \\
    & $(1s2p_{3/2})_1$ &   -0.05x93(1) &    0.0x025     &   -0.01x56(1)  &   -0.07x25(1) \\
    & $(1s2p_{3/2})_2$ &   -0.10x06(1) &    0.0x088     &    0.00x31(1)  &   -0.08x87(1) \\
    &   off-diag.      &    0.03x06    &   -0.0x050     &    0.00x39(1)  &    0.02x94(1) \\
\colrule\hline
\end{longtable*}
\end{ruledtabular}
\endgroup

%%%%%%%%%%%%%%%%%%%%%%%%%%%%%%%%%%%%%%%%%%%%%%%%%%%%%%%%%%%%%%%%%%%%%%%%
%
%
%
%%%%%%%%%%%%%%%%%%%%%%%%%%%%%%%%%%%%%%%%%%%%%%%%%%%%%%%%%%%%%%%%%%%%%%%%
\begingroup
\begin{table}
\caption{Right-hand-side (r.h.s.) and left-hand-side (l.h.s.) of
Eq.~(\ref{id2}) (for the screened self-energy correction) and those of
Eq.~(\ref{id1}) (for the screened vacuum-polarization and two-photon exchange
corrections), in eV. The comparison is valid to the leading order in $\aZ$
only. The last column demonstrates that the difference (r.h.s.-l.h.s.) for
the two-photon exchange correction arises predominantly from effects to order
$\alpha^2(\aZ)^4$.
 \label{tab:identity}}
\begin{ruledtabular}
\begin{tabular}{c..|..|...}
%\hline\hline
 $Z$   & \multicolumn{2}{c}{Scr.SE} &
                \multicolumn{2}{c}{Scr.VP} &
                     \multicolumn{3}{c}{2-ph.exch.} \\
       & \multicolumn{1}{c}{l.h.s.} & \multicolumn{1}{c}{r.h.s.} & \multicolumn{1}{c}{l.h.s.} & \multicolumn{1}{c}{r.h.s.}  & \multicolumn{1}{c}{l.h.s.} & \multicolumn{1}{c}{r.h.s.} & \multicolumn{1}{c}{(r.h.s.-l.h.s.)$/(\aZ)^4$}  \\
\hline
 12 &     0.0x006 &  0.0x006 &     0.0x001 &  0.0x001 &    -0.0x003 & -0.0x001 &  3.x \\
 14 &     0.0x010 &  0.0x009 &     0.0x001 &  0.0x001 &    -0.0x005 & -0.0x002 &  3.x \\
 16 &     0.0x014 &  0.0x013 &     0.0x002 &  0.0x002 &    -0.0x008 & -0.0x003 &  3.x0 \\
 18 &     0.0x020 &  0.0x019 &     0.0x002 &  0.0x002 &    -0.0x012 & -0.0x004 &  3.x0 \\
 20 &     0.0x027 &  0.0x026 &     0.0x003 &  0.0x003 &    -0.0x018 & -0.0x005 &  3.x0 \\
 30 &     0.0x083 &  0.0x087 &     0.0x012 &  0.0x010 &    -0.0x083 & -0.0x013 &  3.x0 \\
 40 &     0.0x167 &  0.0x205 &     0.0x035 &  0.0x025 &    -0.0x256 & -0.0x025 &  3.x18 \\
%\hline\hline
\end{tabular}
\end{ruledtabular}
\end{table}

%%%%%%%%%%%%%%%%%%%%%%%%%%%%%%%%%%%%%%%%%%%%%%%%%%%%%%%%%%%%%%%%%%%%
%
%
%
%%%%%%%%%%%%%%%%%%%%%%%%%%%%%%%%%%%%%%%%%%%%%%%%%%%%%%%%%%%%%%%%%%%%
\begingroup
\squeezetable
\begin{ruledtabular}
\begin{longtable*}{rl.......}
\caption{Individual contributions to the ionization energies of He-like ions (with the
opposite sign), in eV. For mixing configurations, contributions to the matrix
elements are listed. \label{tab:He:contrib} }
 \\
\colrule\hline
 $Z$& State    & \multicolumn{1}{c}{$\Delta E_{\rm Dirac}$}
                                       & \multicolumn{1}{c}{$\Delta E_{\rm int}$}
                                                    & \multicolumn{1}{c}{$\Delta E^{\rm QED}_{\rm 1el}$}
                                                                      &  \multicolumn{1}{c}{$\Delta E^{\rm QED}_{\rm 2el}$}
                                                                                      & \multicolumn{1}{c}{$\Delta E^{\rm QED}_{\rm h.o.}$}
                                                                                                        & \multicolumn{1}{c}{$\Delta E_{\rm rec}$}
                                                                                                                  & \multicolumn{1}{c}{Total}   \\
\hline
 12 & $(1s)^2$         &     -1962.x9887     &  200.x8973 &    0.x2801    &  -0.0x353(1)  &   0.0x008     & 0.0x412 &     -1761.x8045(1) \\
    & $(1s2s)_0$       &      -490.x9832     &   72.x9751 &    0.x0371    &  -0.0x080(1)  &   0.0x005     & 0.0x096 &      -417.x9688(1) \\
    & $(1s2s)_1$       &      -490.x9832     &   60.x2485 &    0.x0371    &  -0.0x053(1)  &   0.0x001     & 0.0x099 &      -430.x6928(1) \\
    & $(1s2p_{1/2})_0$ &      -490.x9834     &   72.x1736 &   -0.x0010    &  -0.0x024(1)  &   0.0x002     & 0.0x037 &      -418.x8092(1) \\
    & $(1s2p_{1/2})_1$ &      -490.x9834     &   74.x9743 &   -0.x0010    &  -0.0x014(1)  &   0.0x002     & 0.0x075 &      -416.x0038(1) \\
    & $(1s2p_{3/2})_1$ &      -490.x0399     &   77.x7699 &    0.x0012    &  -0.0x010(1)  &   0.0x001     & 0.0x113 &      -412.x2584(1) \\
    & $(1s2p_{3/2})_2$ &      -490.x0399     &   71.x7742 &    0.x0012    &  -0.0x024(1)  &   0.0x003     & 0.0x036 &      -418.x2630(1) \\
    &   off-diag.      &          0x         &    4.x1676 &     0x        &   0.0x010(1)  &  -0.0x001     & 0.0x054 &         4.x1739(1) \\
   \hline
 14 & $(1s)^2$         &     -2673.x7078     &  235.x5743 &    0.x4778    &  -0.0x516(1)  &   0.0x008     & 0.0x487 &     -2437.x6577(1) \\
    & $(1s2s)_0$       &      -668.x8650     &   85.x8200 &    0.x0637    &  -0.0x119(1)  &   0.0x006     & 0.0x115 &      -582.x9812(1) \\
    & $(1s2s)_1$       &      -668.x8650     &   70.x5889 &    0.x0637    &  -0.0x079(1)  &   0.0x002     & 0.0x118 &      -598.x2083(1) \\
    & $(1s2p_{1/2})_0$ &      -668.x8654     &   84.x7685 &   -0.x0017    &  -0.0x037(1)  &   0.0x002(1)  & 0.0x044 &      -584.x0977(1) \\
    & $(1s2p_{1/2})_1$ &      -668.x8654     &   88.x0659 &   -0.x0017    &  -0.0x021(1)  &   0.0x002     & 0.0x090 &      -580.x7941(1) \\
    & $(1s2p_{3/2})_1$ &      -667.x1144     &   91.x3604 &    0.x0022    &  -0.0x015(1)  &   0.0x001     & 0.0x135 &      -575.x7397(1) \\
    & $(1s2p_{3/2})_2$ &      -667.x1144     &   84.x1211 &    0.x0022    &  -0.0x036(1)  &   0.0x004     & 0.0x043 &      -582.x9899(1) \\
    &   off-diag.      &          0x         &    5.x0013 &     0x        &   0.0x015(1)  &  -0.0x002     & 0.0x065 &         5.x0091(1) \\
   \hline
 16 & $(1s)^2$         &     -3495.x0043     &  270.x4823 &    0.x7562    &  -0.0x715(1)  &   0.0x009     & 0.0x563 &     -3223.x7801(1) \\
    & $(1s2s)_0$       &      -874.x5000     &   98.x7466 &    0.x1014    &  -0.0x168(1)  &   0.0x007     & 0.0x134 &      -775.x6547(1) \\
    & $(1s2s)_1$       &      -874.x5000     &   80.x9665 &    0.x1014    &  -0.0x110(1)  &   0.0x002     & 0.0x137 &      -793.x4291(1) \\
    & $(1s2p_{1/2})_0$ &      -874.x5006     &   97.x4677 &   -0.x0028    &  -0.0x056(1)  &   0.0x003(2)  & 0.0x051 &      -777.x0359(2) \\
    & $(1s2p_{1/2})_1$ &      -874.x5006     &  101.x2216 &   -0.x0028    &  -0.0x030(1)  &   0.0x003     & 0.0x105 &      -773.x2742(1) \\
    & $(1s2p_{3/2})_1$ &      -871.x5074     &  104.x9766 &    0.x0038    &  -0.0x022(1)  &   0.0x001     & 0.0x158 &      -766.x5134(1) \\
    & $(1s2p_{3/2})_2$ &      -871.x5074     &   96.x4857 &    0.x0038    &  -0.0x051(1)  &   0.0x005     & 0.0x051 &      -775.x0175(1) \\
    &   off-diag.      &          0x         &    5.x8246 &     0x        &   0.0x021(1)  &  -0.0x002     & 0.0x076 &         5.x8341(1) \\
   \hline
 18 & $(1s)^2$         &     -4427.x4152(1)  &  305.x6561 &    1.x1310(1) &  -0.0x953(1)  &   0.0x009     & 0.0x575 &     -4120.x6651(2) \\
    & $(1s2s)_0$       &     -1108.x0563     &  111.x7675 &    0.x1525    &  -0.0x228(1)  &   0.0x007(1)  & 0.0x138 &      -996.x1445(1) \\
    & $(1s2s)_1$       &     -1108.x0563     &   91.x3873 &    0.x1525    &  -0.0x148(1)  &   0.0x003     & 0.0x141 &     -1016.x5169(1) \\
    & $(1s2p_{1/2})_0$ &     -1108.x0574     &  110.x2884 &   -0.x0043    &  -0.0x080(1)  &   0.0x003(3)  & 0.0x053 &      -997.x7757(3) \\
    & $(1s2p_{1/2})_1$ &     -1108.x0574     &  114.x4515 &   -0.x0043    &  -0.0x042(1)  &   0.0x003     & 0.0x108 &      -993.x6035(1) \\
    & $(1s2p_{3/2})_1$ &     -1103.x2520     &  118.x6221 &    0.x0062    &  -0.0x031(1)  &   0.0x001     & 0.0x162 &      -984.x6105(1) \\
    & $(1s2p_{3/2})_2$ &     -1103.x2520     &  108.x8712 &    0.x0062    &  -0.0x068(1)  &   0.0x005     & 0.0x052 &      -994.x3756(1) \\
    &   off-diag.      &          0x         &    6.x6353 &     0x        &   0.0x028(1)  &  -0.0x002     & 0.0x078 &         6.x6456(1) \\
   \hline
 20 & $(1s)^2$         &     -5471.x5558(2)  &  341.x1317 &    1.x6179(2) &  -0.1x231(1)  &   0.0x008     & 0.0x715 &     -5128.x8570(3) \\
    & $(1s2s)_0$       &     -1369.x7265     &  124.x8955 &    0.x2195(1) &  -0.0x300(1)  &   0.0x008(1)  & 0.0x172 &     -1244.x6235(2) \\
    & $(1s2s)_1$       &     -1369.x7265     &  101.x8571 &    0.x2195(1) &  -0.0x192(1)  &   0.0x003     & 0.0x175 &     -1267.x6513(1) \\
    & $(1s2p_{1/2})_0$ &     -1369.x7284     &  123.x2478 &   -0.x0063    &  -0.0x111(1)  &   0.0x004(4)  & 0.0x066 &     -1246.x4910(4) \\
    & $(1s2p_{1/2})_1$ &     -1369.x7284     &  127.x7658 &   -0.x0063    &  -0.0x057(1)  &   0.0x003(1)  & 0.0x135 &     -1241.x9608(1) \\
    & $(1s2p_{3/2})_1$ &     -1362.x3853     &  132.x3004 &    0.x0097    &  -0.0x042(1)  &   0.0x001     & 0.0x203 &     -1230.x0590(1) \\
    & $(1s2p_{3/2})_2$ &     -1362.x3853     &  121.x2809 &    0.x0097    &  -0.0x089(1)  &   0.0x006     & 0.0x065 &     -1241.x0965(1) \\
    &   off-diag.      &          0x         &    7.x4312 &     0x        &   0.0x036(1)  &  -0.0x003     & 0.0x097 &         7.x4442(1) \\
   \hline
 30 & $(1s)^2$         &    -12395.x3519(21) &  524.x3345 &    6.x305(2)  &  -0.3x292(1)  &  -0.0x002     & 0.1x036 &    -11864.x9380(26) \\
    & $(1s2s)_0$       &     -3108.x3049(2)  &  192.x6165 &    0.x883(1)  &  -0.0x884(1)  &   0.0x010(4)  & 0.0x254 &     -2914.x8679(15) \\
    & $(1s2s)_1$       &     -3108.x3049(2)  &  155.x1570 &    0.x883(1)  &  -0.0x527(1)  &   0.0x005     & 0.0x257 &     -2952.x2919(14) \\
    & $(1s2p_{1/2})_0$ &     -3108.x3193     &  190.x7532 &   -0.x0229(3) &  -0.0x433(1)  &   0.0x006(15) & 0.0x099 &     -2917.x6217(15) \\
    & $(1s2p_{1/2})_1$ &     -3108.x3193     &  195.x9801 &   -0.x0229(3) &  -0.0x195(1)  &   0.0x006(5)  & 0.0x199 &     -2912.x3612(6) \\
    & $(1s2p_{3/2})_1$ &     -3070.x5057     &  201.x3174 &    0.x0546(3) &  -0.0x140(1)  &   0.0x001     & 0.0x295 &     -2869.x1181(4) \\
    & $(1s2p_{3/2})_2$ &     -3070.x5057     &  183.x7941 &    0.x0546(3) &  -0.0x245(1)  &   0.0x012     & 0.0x096 &     -2886.x6708(4) \\
    &   off-diag.      &          0x         &   11.x1222 &     0x        &   0.0x093(1)  &  -0.0x006     & 0.0x140 &        11.x1450(1) \\
   \hline
 40 & $(1s)^2$         &    -22253.x1573(98) &  720.x9169 &   16.x315(5)  &  -0.6x721(1)  &  -0.0x030(12) & 0.1x354 &    -21516.x466(11) \\
    & $(1s2s)_0$       &     -5593.x9685(13) &  265.x1796 &    2.x365(7)  &  -0.2x002(1)  &   0.0x007(5)  & 0.0x334 &     -5326.x5902(69) \\
    & $(1s2s)_1$       &     -5593.x9685(13) &  210.x6536 &    2.x365(7)  &  -0.1x091(1)  &   0.0x008     & 0.0x338 &     -5381.x0246(69) \\
    & $(1s2p_{1/2})_0$ &     -5594.x0369     &  264.x5843 &   -0.x040(2)  &  -0.1x234(1)  &   0.0x008(37) & 0.0x134 &     -5329.x6015(40) \\
    & $(1s2p_{1/2})_1$ &     -5594.x0369     &  268.x0222 &   -0.x040(2)  &  -0.0x510(1)  &   0.0x007(12) & 0.0x260 &     -5326.x0786(20) \\
    & $(1s2p_{3/2})_1$ &     -5471.x5704     &  271.x7376 &    0.x193(2)  &  -0.0x345(1)  &   0.0x001     & 0.0x378 &     -5199.x6369(16) \\
    & $(1s2p_{3/2})_2$ &     -5471.x5704     &  247.x3427 &    0.x193(2)  &  -0.0x503(1)  &   0.0x018     & 0.0x125 &     -5224.x0712(16) \\
    &   off-diag.      &          0x         &   14.x1717 &     0x        &   0.0x179(1)  &  -0.0x010(1)  & 0.0x178 &        14.x2065(1) \\
   \hline
 50 & $(1s)^2$         &    -35226.x611(37)  &  936.x5564 &   33.x961(8)  &  -1.2x016(1)  &  -0.0x077(50) & 0.1x659 &    -34257.x137(38) \\
    & $(1s2s)_0$       &     -8884.x0997(51) &  344.x7875 &    5.x118(22) &  -0.3x966(1)  &  -0.0x001     & 0.0x412 &     -8534.x550(22) \\
    & $(1s2s)_1$       &     -8884.x0997(51) &  269.x3108 &    5.x118(22) &  -0.1x953(1)  &   0.0x011(1)  & 0.0x415 &     -8609.x824(22) \\
    & $(1s2p_{1/2})_0$ &     -8884.x3678(1)  &  347.x6605 &   -0.x006(5)  &  -0.2x936(1)  &   0.0x009(77) & 0.0x170 &     -8536.x9893(91) \\
    & $(1s2p_{1/2})_1$ &     -8884.x3678(1)  &  345.x6503 &   -0.x006(5)  &  -0.1x152(1)  &   0.0x009(26) & 0.0x315 &     -8538.x8067(55) \\
    & $(1s2p_{3/2})_1$ &     -8575.x5139     &  344.x0793 &    0.x523(5)  &  -0.0x725(1)  &  -0.0x002(2)  & 0.0x447 &     -8230.x9398(49) \\
    & $(1s2p_{3/2})_2$ &     -8575.x5139     &  312.x2766 &    0.x523(5)  &  -0.0x887(1)  &   0.0x025(1)  & 0.0x152 &     -8262.x7855(49) \\
    &   off-diag.      &          0x         &   16.x3734 &     0x        &   0.0x294(1)  &  -0.0x015(3)  & 0.0x206 &        16.x4220(3) \\
   \hline
 60 & $(1s)^2$         &    -51577.x89(11)   & 1178.x1908 &   61.x92(2)   &  -1.9x930(2)  &  -0.0x14(15)  & 0.2x152 &    -50339.x58(12) \\
    & $(1s2s)_0$       &    -13062.x076(17)  &  434.x2620 &    9.x74(5)   &  -0.7x258(2)  &  -0.0x014(24) & 0.0x537 &    -12618.x746(55) \\
    & $(1s2s)_1$       &    -13062.x076(17)  &  332.x3354 &    9.x74(5)   &  -0.3x219(2)  &   0.0x015(3)  & 0.0x541 &    -12720.x265(55) \\
    & $(1s2p_{1/2})_0$ &    -13062.x9663(6)  &  443.x7641 &    0.x20(1)   &  -0.6x203(1)  &   0.0x01(14)  & 0.0x227 &    -12619.x596(19) \\
    & $(1s2p_{1/2})_1$ &    -13062.x9663(6)  &  431.x1435 &    0.x20(1)   &  -0.2x367(1)  &   0.0x010(46) & 0.0x400 &    -12631.x816(13) \\
    & $(1s2p_{3/2})_1$ &    -12395.x4629     &  418.x8977 &    1.x20(1)   &  -0.1x366(2)  &  -0.0x005(9)  & 0.0x546 &    -11975.x449(12) \\
    & $(1s2p_{3/2})_2$ &    -12395.x4629     &  378.x9465 &    1.x20(1)   &  -0.1x426(2)  &   0.0x034(1)  & 0.0x193 &    -12015.x437(12) \\
    &   off-diag.      &          0x         &   17.x5341 &     0x        &   0.0x437(2)  &  -0.0x021(5)  & 0.0x246 &        17.x6003(6) \\
   \hline
 70 & $(1s)^2$         &    -71678.x25(34)   & 1454.x7177 &  103.x45( 5)  &  -3.1x581(2)  &  -0.0x23(36)  & 0.2x612 &    -70123.x00(34) \\
    & $(1s2s)_0$       &    -18247.x262(53)  &  537.x4108 &   17.x07(10)  &  -1.2x610(2)  &  -0.0x033(84) & 0.0x658 &    -17693.x98(11) \\
    & $(1s2s)_1$       &    -18247.x262(53)  &  401.x3107 &   17.x07(10)  &  -0.5x043(2)  &   0.0x019(5)  & 0.0x662 &    -17829.x31(11) \\
    & $(1s2p_{1/2})_0$ &    -18250.x1817(30) &  558.x0871 &    0.x84( 3)  &  -1.2x094(2)  &   0.0x01(25)  & 0.0x283 &    -17692.x431(37) \\
    & $(1s2p_{1/2})_1$ &    -18250.x1817(30) &  527.x6354 &    0.x84( 3)  &  -0.4x550(2)  &   0.0x011(74) & 0.0x469 &    -17722.x109(28) \\
    & $(1s2p_{3/2})_1$ &    -16948.x0254     &  496.x7905 &    2.x44( 3)  &  -0.2x388(2)  &  -0.0x010(24) & 0.0x608 &    -16448.x971(27) \\
    & $(1s2p_{3/2})_2$ &    -16948.x0254     &  447.x7121 &    2.x44( 3)  &  -0.2x158(2)  &   0.0x043(1)  & 0.0x225 &    -16498.x059(27) \\
    &   off-diag.      &          0x         &   17.x4750 &     0x        &   0.0x603(2)  &  -0.0x030(9)  & 0.0x266 &        17.x5590(9) \\
   \hline
 80 & $(1s)^2$         &    -96061.x17(14)   & 1778.x3460 &  162.x76(12)  &  -4.8x682(3)  &  -0.0x35(78)  & 0.3x326 &    -94124.x63(20) \\
    & $(1s2s)_0$       &    -24612.x823(24)  &  659.x7043 &   28.x31(16)  &  -2.1x164(2)  &  -0.0x06(21)  & 0.0x853 &    -23926.x84(16) \\
    & $(1s2s)_1$       &    -24612.x823(24)  &  478.x4429 &   28.x31(16)  &  -0.7x652(2)  &   0.0x023(9)  & 0.0x858 &    -24106.x74(16) \\
    & $(1s2p_{1/2})_0$ &    -24621.x4092(19) &  698.x2096 &    2.x41(5)   &  -2.2x345(1)  &   0.0x01(41)  & 0.0x369 &    -23922.x985(66) \\
    & $(1s2p_{1/2})_1$ &    -24621.x4092(19) &  639.x7159 &    2.x41(5)   &  -0.8x367(1)  &   0.0x01(11)  & 0.0x570 &    -23980.x061(53) \\
    & $(1s2p_{3/2})_1$ &    -22253.x6733     &  578.x4003 &    4.x56(5)   &  -0.3x926(3)  &  -0.0x017(52) & 0.0x686 &    -21671.x042(52) \\
    & $(1s2p_{3/2})_2$ &    -22253.x6733     &  518.x9620 &    4.x56(5)   &  -0.3x132(3)  &   0.0x054(1)  & 0.0x267 &    -21730.x436(52) \\
    &   off-diag.      &          0x        &   16.x0290 &     0 x        &   0.0x789(2)  &  -0.0x040(15) & 0.0x289 &        16.x133(2) \\
   \hline
 83 & $(1s)^2$         &   -104318.x14(45)   & 1887.x0340 &  184.x80(15)  &  -5.5x216(7)  &  -0.0x38(96)  & 0.3x620 &   -102251.x50(48) \\
    & $(1s2s)_0$       &    -26787.x971(79)  &  701.x2539 &   32.x69(18)  &  -2.4x602(2)  &  -0.0x07(27)  & 0.0x938 &    -26056.x40(20) \\
    & $(1s2s)_1$       &    -26787.x971(79)  &  503.x6141 &   32.x69(18)  &  -0.8x635(2)  &   0.0x025(10) & 0.0x943 &    -26252.x43(20) \\
    & $(1s2p_{1/2})_0$ &    -26799.x9095(69) &  746.x8793 &    3.x18(6)   &  -2.6x669(1)  &   0.0x01(47)  & 0.0x404 &    -26052.x473(79) \\
    & $(1s2p_{1/2})_1$ &    -26799.x9095(69) &  677.x3537 &    3.x18(6)   &  -0.9x985(1)  &   0.0x01(12)  & 0.0x611 &    -26120.x309(64) \\
    & $(1s2p_{3/2})_1$ &    -23995.x8077     &  603.x7089 &    5.x42(6)   &  -0.4x511(4)  &  -0.0x020(63) & 0.0x715 &    -23387.x064(63) \\
    & $(1s2p_{3/2})_2$ &    -23995.x8077     &  540.x8844 &    5.x42(6)   &  -0.3x482(4)  &   0.0x057     & 0.0x283 &    -23449.x822(62) \\
    &   off-diag.      &          0x         &   15.x3022 &     0x        &   0.0x848(4)  &  -0.0x043(17) & 0.0x297 &        15.x412(2) \\
   \hline
 90 & $(1s)^2$         &   -125495.x06(35)   & 2166.x4366 &  245.x28(27)  &  -7.3x73(1)   &  -0.0x5(16)   & 0.4x338 &   -123090.x46(47) \\
    & $(1s2s)_0$       &    -32413.x922(65)  &  809.x3517 &   45.x19(21)  &  -3.4x708(3)  &  -0.0x09(45)  & 0.1x168 &    -31562.x77(23) \\
    & $(1s2s)_1$       &    -32413.x922(65)  &  566.x9102 &   45.x19(21)  &  -1.1x376(2)  &   0.0x029(14) & 0.1x174 &    -31802.x86(22) \\
    & $(1s2p_{1/2})_0$ &    -32440.x5502(70) &  875.x9130 &    5.x82(9)   &  -3.9x928(2)  &   0.0x02(66)  & 0.0x496 &    -31562.x76(11) \\
    & $(1s2p_{1/2})_1$ &    -32440.x5502(70) &  774.x5561 &    5.x82(9)   &  -1.4x974(1)  &   0.0x01(16)  & 0.0x707 &    -31661.x599(95) \\
    & $(1s2p_{3/2})_1$ &    -28337.x2409     &  664.x4053 &    7.x93(9)   &  -0.6x138(5)  &  -0.0x026(97) & 0.0x759 &    -27665.x448(94) \\
    & $(1s2p_{3/2})_2$ &    -28337.x2409     &  593.x1280 &    7.x93(9)   &  -0.4x420(5)  &   0.0x066     & 0.0x314 &    -27736.x589(93) \\
    &   off-diag.      &          0x         &   13.x0506 &     0x        &   0.0x991(4)  &  -0.0x052(23) & 0.0x305 &        13.x173(2) \\
   \hline
 92 & $(1s)^2$         &   -132081.x13(40)   & 2253.x9270 &  265.x16(33)  &  -8.0x02(2)   &  -0.0x5(18)   & 0.4x600 &   -129569.x84(55) \\
    & $(1s2s)_0$       &    -34177.x718(76)  &  843.x6057 &   49.x44(22)  &  -3.8x259(4)  &  -0.0x09(51)  & 0.1x260 &    -33288.x42(24) \\
    & $(1s2s)_1$       &    -34177.x718(76)  &  586.x3549 &   49.x44(22)  &  -1.2x301(3)  &   0.0x030(16) & 0.1x266 &    -33543.x06(23) \\
    & $(1s2p_{1/2})_0$ &    -34211.x0649(86) &  917.x4965 &    6.x86(10)  &  -4.4x740(3)  &   0.0x02(73)  & 0.0x531 &    -33291.x13(13) \\
    & $(1s2p_{1/2})_1$ &    -34211.x0649(86) &  805.x1933 &    6.x86(10)  &  -1.6x798(2)  &   0.0x01(17)  & 0.0x743 &    -33400.x62(11) \\
    & $(1s2p_{3/2})_1$ &    -29649.x8340     &  682.x1947 &    8.x80(10)  &  -0.6x677(5)  &  -0.0x03(11)  & 0.0x774 &    -28959.x44(10) \\
    & $(1s2p_{3/2})_2$ &    -29649.x8340     &  608.x3558 &    8.x80(10)  &  -0.4x725(5)  &   0.0x068     & 0.0x324 &    -29033.x12(10) \\
    &   off-diag.      &          0x         &   12.x2592 &     0x        &   0.1x031(4)  &  -0.0x054(25) & 0.0x308 &        12.x383(3) \\
   \hline
100 & $(1s)^2$         &   -161165.x5(6.0)   & 2646.x5635 &  358.x30(63)  & -11.0x79(4)   &  -0.0x6(30)   & 0.6x180 &   -158171.x1(6.1) \\
    & $(1s2s)_0$       &    -42048.x7(1.2)   &  999.x8620 &   70.x19(20)  &  -5.6x271(8)  &  -0.0x12(86)  & 0.1x895 &    -40984.x1(1.3) \\
    & $(1s2s)_1$       &    -42048.x7(1.2)   &  671.x7243 &   70.x19(20)  &  -1.6x730(5)  &   0.0x035(23) & 0.1x902 &    -41308.x3(1.3) \\
    & $(1s2p_{1/2})_0$ &    -42127.x25(19)   & 1111.x1289 &   12.x82(16)  &  -7.0x293(6)  &   0.0x0(11)   & 0.0x759 &    -41010.x25(27) \\
    & $(1s2p_{1/2})_1$ &    -42127.x25(19)   &  944.x3801 &   12.x82(16)  &  -2.6x547(6)  &   0.0x01(21)  & 0.0x984 &    -41172.x60(25) \\
    & $(1s2p_{3/2})_1$ &    -35228.x5685     &  755.x4926 &   13.x05(16)  &  -0.9x195(7)  &  -0.0x04(16)  & 0.0x866 &    -34460.x87(16) \\
    & $(1s2p_{3/2})_2$ &    -35228.x5685     &  670.x7584 &   13.x05(16)  &  -0.6x128(7)  &   0.0x079(2)  & 0.0x382 &    -34545.x33(16) \\
    &   off-diag.      &          0x         &    8.x4030 &     0x        &   0.1x202(4)  &  -0.0x066(34) & 0.0x328 &         8.x550(3) \\
   \hline
\colrule\hline
\end{longtable*}
\end{ruledtabular}
\endgroup

\begingroup
\squeezetable
\begin{ruledtabular}
%\begin{longtable*}{rcccccccc}
\begin{longtable*}{rc.......}
\caption{Total ionization energies (in eV) for $n=1$ and $n=2$ states of
He-like ions. ``RMS" denotes the root-mean-square radii expressed in Fermi.
\label{tab:total}}
 \\
\colrule\hline
 Z  &  RMS  &      \multicolumn{1}{c}{$1\,^1S_0$}
                                 &       \multicolumn{1}{c}{$2\,^1S_0$}
                                                      &      \multicolumn{1}{c}{ $2\,^3S_1$}
                                                                           &       \multicolumn{1}{c}{ $2\,^3P_0$}
                                                                                                &      \multicolumn{1}{c}{ $2\,^3P_1$}
                                                                                                                     &      \multicolumn{1}{c}{ $2\,^1P_1$}
                                                                                                                                         &    \multicolumn{1}{c}{  $2\,^3P_2$ }   \\
\hline
 12 & 3.057 &  1761.x8045(1)     &   417.x9688(1)     &   430.x6928(1)     &   418.x8092(1)     &   418.x7058(1)     &   409.x5564(1)     &   418.x2630(1)  \\
 13 & 3.063 &  2085.x9766(1)     &   497.x0264(1)     &   510.x9969(1)     &   498.x0059(2)     &   497.x8513(1)     &   487.x6853(1)     &   497.x2157(1)  \\
 14 & 3.123 &  2437.x6577(1)     &   582.x9812(1)     &   598.x2083(1)     &   584.x0977(2)     &   583.x8774(1)     &   572.x6564(1)     &   582.x9899(1)  \\
 15 & 3.190 &  2816.x9083(1)     &   675.x8517(1)     &   692.x3465(1)     &   677.x1018(2)     &   676.x8002(1)     &   664.x4774(1)     &   675.x5896(1)  \\
 16 & 3.263 &  3223.x7801(2)     &   775.x6547(1)     &   793.x4291(1)     &   777.x0359(2)     &   776.x6364(1)     &   763.x1511(1)     &   775.x0175(1)  \\
 17 & 3.388 &  3658.x3431(2)     &   882.x4119(2)     &   901.x4785(1)     &   883.x9201(3)     &   883.x4061(1)     &   868.x6849(1)     &   881.x2782(1)  \\
 18 & 3.427 &  4120.x6651(3)     &   996.x1445(2)     &   1016.x5169(1)    &   997.x7757(3)     &   997.x1309(1)     &   981.x0831(1)     &   994.x3756(1)  \\
 19 & 3.435 &  4610.x8065(3)     &   1116.x8726(2)    &   1138.x5651(2)    &   1118.x6242(4)    &   1117.x8332(2)    &   1100.x3452(1)    &   1114.x3132(1)  \\
 20 & 3.478 &  5128.x8570(4)     &   1244.x6235(2)    &   1267.x6513(2)    &   1246.x4910(4)    &   1245.x5403(2)    &   1226.x4795(1)    &   1241.x0965(1)  \\
 21 & 3.546 &  5674.x9027(5)     &   1379.x4240(3)    &   1403.x8033(2)    &   1381.x4021(5)    &   1380.x2810(2)    &   1359.x4906(1)    &   1374.x7310(1)  \\
 22 & 3.592 &  6249.x0215(6)     &   1521.x2993(3)    &   1547.x0472(3)    &   1523.x3840(6)    &   1522.x0845(2)    &   1499.x3776(1)    &   1515.x2210(1)  \\
 23 & 3.600 &  6851.x3098(7)     &   1670.x2794(4)    &   1697.x4139(3)    &   1672.x4660(7)    &   1670.x9837(2)    &   1646.x1447(2)    &   1662.x5723(1)  \\
 24 & 3.645 &  7481.x8615(9)     &   1826.x3943(5)    &   1854.x9342(4)    &   1828.x6783(8)    &   1827.x0127(3)    &   1799.x7935(2)    &   1816.x7904(2)  \\
 25 & 3.706 &  8140.x7858(11)    &   1989.x6779(6)    &   2019.x6431(5)    &   1992.x0538(9)    &   1990.x2085(3)    &   1960.x3290(2)    &   1977.x8820(2)  \\
 26 & 3.738 &  8828.x1864(13)    &   2160.x1629(7)    &   2191.x5742(7)    &   2162.x6259(10)   &   2160.x6082(4)    &   2127.x7522(2)    &   2145.x8529(2)  \\
 27 & 3.788 &  9544.x1817(15)    &   2337.x8865(9)    &   2370.x7658(8)    &   2340.x4307(11)   &   2338.x2522(4)    &   2302.x0688(2)    &   2320.x7104(3)  \\
 28 & 3.776 &  10288.x8845(18)   &   2522.x8843(11)   &   2557.x2543(10)   &   2525.x5046(12)   &   2523.x1803(5)    &   2483.x2797(3)    &   2502.x4605(3)  \\
 29 & 3.883 &  11062.x4295(22)   &   2715.x1988(13)   &   2751.x0833(12)   &   2717.x8885(13)   &   2715.x4371(5)    &   2671.x3951(3)    &   2691.x1119(3)  \\
 30 & 3.928 &  11864.x9380(26)   &   2914.x8679(15)   &   2952.x2919(14)   &   2917.x6217(15)   &   2915.x0645(6)    &   2866.x4147(4)    &   2886.x6708(4)  \\
 31 & 3.996 &  12696.x5555(31)   &   3121.x9372(18)   &   3160.x9268(17)   &   3124.x7480(17)   &   3122.x1099(7)    &   3068.x3488(5)    &   3089.x1461(5)  \\
 32 & 4.072 &  13557.x4188(37)   &   3336.x4503(21)   &   3377.x0324(20)   &   3339.x3118(19)   &   3336.x6195(8)    &   3277.x2016(5)    &   3298.x5453(6)  \\
 33 & 4.096 &  14447.x6761(44)   &   3558.x4532(25)   &   3600.x6561(24)   &   3561.x3593(21)   &   3558.x6415(9)    &   3492.x9786(6)    &   3514.x8764(6)  \\
 34 & 4.140 &  15367.x4889(51)   &   3787.x9973(29)   &   3831.x8502(29)   &   3790.x9401(23)   &   3788.x2280(10)   &   3715.x6912(7)    &   3738.x1490(7)  \\
 35 & 4.163 &  16317.x0085(59)   &   4025.x1296(34)   &   4070.x6630(34)   &   4028.x1037(25)   &   4025.x4293(12)   &   3945.x3431(8)    &   3968.x3710(9)  \\
 36 & 4.188 &  17296.x4182(68)   &   4269.x9080(39)   &   4317.x1534(39)   &   4272.x9045(28)   &   4270.x3024(13)   &   4181.x9484(10)   &   4205.x5532(10)  \\
 37 & 4.204 &  18305.x8805(77)   &   4522.x3840(46)   &   4571.x3742(45)   &   4525.x3960(31)   &   4522.x9014(14)   &   4425.x5118(11)   &   4449.x7039(11)  \\
 38 & 4.224 &  19345.x5841(89)   &   4782.x6167(53)   &   4833.x3860(52)   &   4785.x6363(34)   &   4783.x2861(16)   &   4676.x0457(13)   &   4700.x8338(13)  \\
 39 & 4.243 &  20415.x713(10)    &   5050.x6647(61)   &   5103.x2482(60)   &   5053.x6841(37)   &   5051.x5162(18)   &   4933.x5581(14)   &   4958.x9527(14)  \\
 40 & 4.270 &  21516.x465(11)    &   5326.x5902(69)   &   5381.x0246(69)   &   5329.x6015(40)   &   5327.x6551(20)   &   5198.x0604(15)   &   5224.x0712(16)  \\
 41 & 4.324 &  22648.x042(12)    &   5610.x4573(79)   &   5666.x7809(79)   &   5613.x4525(44)   &   5611.x7684(22)   &   5469.x5642(18)   &   5496.x2005(18)  \\
 42 & 4.407 &  23810.x651(14)    &   5902.x3327(90)   &   5960.x5848(89)   &   5905.x3040(48)   &   5903.x9240(25)   &   5748.x0806(20)   &   5775.x3519(20)  \\
 43 & 4.424 &  25004.x529(16)    &   6202.x286(10)    &   6262.x508(10)    &   6205.x2242(52)   &   6204.x1906(27)   &   6033.x6179(22)   &   6061.x5357(23)  \\
 44 & 4.481 &  26229.x891(18)    &   6510.x390(11)    &   6572.x623(11)    &   6513.x2861(56)   &   6512.x6431(31)   &   6326.x1919(25)   &   6354.x7651(25)  \\
 45 & 4.494 &  27486.x978(20)    &   6826.x720(12)    &   6891.x008(12)    &   6829.x5632(61)   &   6829.x3555(34)   &   6625.x8116(28)   &   6655.x0511(29)  \\
 46 & 4.532 &  28776.x030(23)    &   7151.x351(14)    &   7217.x739(14)    &   7154.x1334(66)   &   7154.x4075(37)   &   6932.x4916(32)   &   6962.x4069(32)  \\
 47 & 4.544 &  30097.x313(26)    &   7484.x365(16)    &   7552.x901(16)    &   7487.x0762(72)   &   7487.x8795(41)   &   7246.x2430(35)   &   7276.x8447(36)  \\
 48 & 4.610 &  31451.x058(30)    &   7825.x844(18)    &   7896.x575(18)    &   7828.x4758(78)   &   7829.x8578(46)   &   7567.x0820(40)   &   7598.x3787(40)  \\
 49 & 4.614 &  32837.x588(33)    &   8175.x877(20)    &   8248.x853(20)    &   8178.x4166(84)   &   8180.x4270(50)   &   7895.x0172(44)   &   7927.x0206(44)  \\
 50 & 4.655 &  34257.x137(37)    &   8534.x550(22)    &   8609.x824(22)    &   8536.x9893(91)   &   8539.x6802(55)   &   8230.x0663(49)   &   8262.x7855(49)  \\
 51 & 4.681 &  35710.x021(42)    &   8901.x956(24)    &   8979.x581(24)    &   8904.x2848(98)   &   8907.x7092(60)   &   8572.x2401(54)   &   8605.x6865(54)  \\
 52 & 4.742 &  37196.x516(48)    &   9278.x191(27)    &   9358.x223(27)    &   9280.x401(10)    &   9284.x6147(66)   &   8921.x5577(59)   &   8955.x7395(59)  \\
 53 & 4.749 &  38716.x991(53)    &   9663.x359(29)    &   9745.x855(29)    &   9665.x435(11)    &   9670.x4932(72)   &   9278.x0263(65)   &   9312.x9568(65)  \\
 54 & 4.787 &  40271.x717(60)    &   10057.x559(32)   &   10142.x577(32)   &   10059.x493(12)   &   10065.x4542(79)  &   9641.x6682(72)   &   9677.x3560(72)  \\
 55 & 4.804 &  41861.x068(67)    &   10460.x900(35)   &   10548.x504(35)   &   10462.x681(13)   &   10469.x6036(86)  &   10012.x4945(79)  &   10048.x9509(79)  \\
 56 & 4.839 &  43485.x358(75)    &   10873.x491(39)   &   10963.x742(39)   &   10875.x110(14)   &   10883.x0560(94)  &   10390.x5238(86)  &   10427.x7586(86)  \\
 57 & 4.855 &  45144.x988(83)    &   11295.x452(42)   &   11388.x416(42)   &   11296.x894(15)   &   11305.x926(10)   &   10775.x7693(94)  &   10813.x7940(94)  \\
 58 & 4.877 &  46840.x299(93)    &   11726.x896(46)   &   11822.x641(46)   &   11728.x153(16)   &   11738.x337(11)   &   11168.x248(10)   &   11207.x074(10)  \\
 59 & 4.892 &  48571.x70(10)     &   12167.x953(50)   &   12266.x549(50)   &   12169.x011(17)   &   12180.x414(12)   &   11567.x978(11)   &   11607.x616(11)  \\
 60 & 4.914 &  50339.x57(11)     &   12618.x745(54)   &   12720.x265(54)   &   12619.x596(18)   &   12632.x287(13)   &   11974.x976(12)   &   12015.x437(12)  \\
 61 & 4.962 &  52144.x28(12)     &   13079.x399(59)   &   13183.x917(59)   &   13080.x041(20)   &   13094.x093(14)   &   12389.x261(13)   &   12430.x554(13)  \\
 62 & 5.092 &  53986.x08(14)     &   13550.x028(64)   &   13657.x622(64)   &   13550.x484(21)   &   13565.x970(15)   &   12810.x849(14)   &   12852.x988(14)  \\
 63 & 5.118 &  55865.x89(16)     &   14030.x843(69)   &   14141.x593(69)   &   14031.x069(23)   &   14048.x066(16)   &   13239.x757(15)   &   13282.x753(15)  \\
 64 & 5.159 &  57783.x89(18)     &   14521.x952(74)   &   14635.x942(74)   &   14521.x947(24)   &   14540.x533(17)   &   13676.x007(17)   &   13719.x870(16)  \\
 65 & 5.099 &  59740.x95(19)     &   15023.x567(79)   &   15140.x883(79)   &   15023.x271(26)   &   15043.x528(19)   &   14119.x614(18)   &   14164.x358(18)  \\
 66 & 5.224 &  61736.x53(22)     &   15535.x693(86)   &   15656.x421(86)   &   15535.x200(28)   &   15557.x210(20)   &   14570.x601(19)   &   14616.x237(19)  \\
 67 & 5.155 &  63772.x63(24)     &   16058.x718(91)   &   16182.x954(91)   &   16057.x906(30)   &   16081.x757(22)   &   15028.x984(21)   &   15075.x525(21)  \\
 68 & 5.250 &  65848.x23(27)     &   16592.x583(98)   &   16720.x419(98)   &   16591.x556(32)   &   16617.x336(24)   &   15494.x786(23)   &   15542.x244(23)  \\
 69 & 5.192 &  67965.x42(29)     &   17137.x69(10)    &   17269.x23(10)    &   17136.x338(34)   &   17164.x141(25)   &   15968.x027(24)   &   16016.x414(24)  \\
 70 & 5.317 &  70122.x99(34)     &   17693.x97(11)    &   17829.x31(11)    &   17692.x430(36)   &   17722.x351(27)   &   16448.x728(26)   &   16498.x058(26)  \\
 71 & 5.246 &  72323.x72(36)     &   18261.x94(12)    &   18401.x19(12)    &   18260.x039(38)   &   18292.x178(29)   &   16936.x909(28)   &   16987.x195(28)  \\
 72 & 5.349 &  74565.x87(41)     &   18841.x44(13)    &   18984.x70(12)    &   18839.x354(41)   &   18873.x814(32)   &   17432.x594(30)   &   17483.x849(30)  \\
 73 & 5.354 &  76851.x98(45)     &   19432.x97(13)    &   19580.x37(13)    &   19430.x599(44)   &   19467.x486(34)   &   17935.x803(33)   &   17988.x041(32)  \\
 74 & 5.373 &  79181.x87(50)     &   20036.x62(14)    &   20188.x27(14)    &   20033.x990(46)   &   20073.x413(36)   &   18446.x561(35)   &   18499.x794(35)  \\
 75 & 5.351 &  81556.x77(54)     &   20652.x69(15)    &   20808.x72(15)    &   20649.x760(49)   &   20691.x835(39)   &   18964.x889(37)   &   19019.x131(37)  \\
 76 & 5.406 &  83976.x26(60)     &   21281.x25(16)    &   21441.x78(16)    &   21278.x139(52)   &   21322.x982(41)   &   19490.x812(40)   &   19546.x078(40)  \\
 77 & 5.401 &  86442.x47(66)     &   21922.x76(18)    &   22087.x92(18)    &   21919.x389(56)   &   21967.x123(44)   &   20024.x352(43)   &   20080.x656(42)  \\
 78 & 5.427 &  88955.x17(73)     &   22577.x32(19)    &   22747.x25(19)    &   22573.x761(59)   &   22624.x513(47)   &   20565.x534(46)   &   20622.x892(45)  \\
 79 & 5.437 &  91515.x78(80)     &   23245.x28(20)    &   23420.x12(20)    &   23241.x533(63)   &   23295.x435(51)   &   21114.x385(49)   &   21172.x810(48)  \\
 80 & 5.467 &  94124.x62(20)     &   23926.x84(16)    &   24106.x74(16)    &   23922.x985(66)   &   23980.x173(53)   &   21670.x929(52)   &   21730.x436(52)  \\
 81 & 5.483 &  96783.x07(98)     &   24622.x37(23)    &   24807.x48(23)    &   24618.x417(71)   &   24679.x035(58)   &   22235.x191(55)   &   22295.x795(55)  \\
 82 & 5.504 &  99491.x78(52)     &   25332.x13(19)    &   25522.x62(19)    &   25328.x139(74)   &   25392.x333(60)   &   22807.x199(59)   &   22868.x915(58)  \\
 83 & 5.533 &  102251.x50(48)    &   26056.x40(19)    &   26252.x43(19)    &   26052.x472(78)   &   26120.x396(64)   &   23386.x977(62)   &   23449.x821(62)  \\
 84 & 5.531 &  105064.x1(1.3)    &   26795.x67(29)    &   26997.x42(29)    &   26791.x770(85)   &   26863.x584(70)   &   23974.x554(66)   &   24038.x542(66)  \\
 85 & 5.539 &  107930.x0(1.4)    &   27550.x15(31)    &   27757.x80(31)    &   27546.x379(90)   &   27622.x249(75)   &   24569.x959(70)   &   24635.x105(70)  \\
 86 & 5.632 &  110847.x9(1.6)    &   28319.x82(34)    &   28533.x55(34)    &   28316.x644(96)   &   28396.x741(80)   &   25173.x222(75)   &   25239.x543(74)  \\
 87 & 5.640 &  113823.x5(1.7)    &   29105.x83(37)    &   29325.x84(37)    &   29103.x01(10)    &   29187.x523(86)   &   25784.x366(79)   &   25851.x878(79)  \\
 88 & 5.662 &  116855.x4(1.9)    &   29908.x12(40)    &   30134.x62(40)    &   29905.x87(10)    &   29994.x978(92)   &   26403.x424(84)   &   26472.x143(83)  \\
 89 & 5.670 &  119945.x7(2.1)    &   30727.x27(44)    &   30960.x47(44)    &   30725.x66(11)    &   30819.x562(99)   &   27030.x425(88)   &   27100.x367(88)  \\
 90 & 5.802 &  123090.x45(46)    &   31562.x76(22)    &   31802.x86(22)    &   31562.x75(11)    &   31661.x642(94)   &   27665.x405(93)   &   27736.x589(93)  \\
 91 & 5.700 &  126304.x7(2.5)    &   32417.x55(52)    &   32664.x83(52)    &   32417.x86(13)    &   32521.x96(11)    &   28308.x381(99)   &   28380.x818(98)  \\
 92 & 5.860 &  129569.x84(54)    &   33288.x42(23)    &   33543.x06(22)    &   33291.x13(12)    &   33400.x65(10)    &   28959.x40(10)    &   29033.x11(10)  \\
 93 & 5.744 &  132910.x9(3.1)    &   34180.x25(63)    &   34442.x58(63)    &   34183.x54(15)    &   34298.x73(13)    &   29618.x48(11)    &   29693.x48(10)  \\
 94 & 5.794 &  136309.x1(3.4)    &   35089.x75(70)    &   35359.x99(69)    &   35095.x26(16)    &   35216.x35(14)    &   30285.x67(11)    &   30361.x97(11)  \\
 95 & 5.787 &  139776.x9(3.7)    &   36019.x54(76)    &   36297.x99(76)    &   36027.x09(17)    &   36154.x35(15)    &   30960.x99(12)    &   31038.x62(12)  \\
 96 & 5.815 &  143310.x9(4.1)    &   36969.x20(84)    &   37256.x14(84)    &   36979.x56(19)    &   37113.x25(16)    &   31644.x49(12)    &   31723.x45(12)  \\
 97 & 5.815 &  146916.x8(4.5)    &   37940.x10(93)    &   38235.x85(92)    &   37953.x42(20)    &   38093.x82(18)    &   32336.x18(13)    &   32416.x50(13)  \\
 98 & 5.843 &  150592.x5(5.0)    &   38932.x1(1.0)    &   39236.x9(1.0)    &   38949.x28(22)    &   39096.x68(20)    &   33036.x12(14)    &   33117.x80(14)  \\
 99 & 5.850 &  154343.x8(5.5)    &   39946.x6(1.1)    &   40261.x0(1.1)    &   39967.x97(24)    &   40122.x68(22)    &   33744.x33(15)    &   33827.x40(14)  \\
100 & 5.857 &  158171.x1(6.0)    &   40984.x1(1.2)    &   41308.x2(1.2)    &   41010.x25(26)    &   41172.x61(24)    &   34460.x85(15)    &   34545.x32(15)  \\
\colrule\hline
\end{longtable*}
\end{ruledtabular}
\endgroup

%%%%%%%%%%%%%%%%%%%%%%%%%%%%%%%%%%%%%%%%%%%%%%%%%%%%%%%%%%%%%%%%%%%%%%%%
%%%%%
%%%%%
%%%%%%%%%%%%%%%%%%%%%%%%%%%%%%%%%%%%%%%%%%%%%%%%%%%%%%%%%%%%%%%%%%%%%%%
\begingroup
\begin{ruledtabular}
\begin{longtable*}{r.....c}
\caption{Comparison of theoretical and experimental transition energies.
Units are cm$^{-1}$ or eV as noted. \label{tab:transen}} \\ \colrule\hline
$Z$ &  \multicolumn{1}{c}{This work}
                   & \multicolumn{1}{c}{Plante {\em et al.} \cite{plante:94}}
                   & \multicolumn{1}{c}{Chen {\em et al.} \cite{chen:93:pra}}
                   & \multicolumn{1}{c}{Johnson {\em et al.} \cite{johnson:92}}
                               & \multicolumn{1}{c}{Experiment}
                                       & \multicolumn{1}{c}{Reference}           \\
\hline
\multicolumn{5}{l}{
$2^3P_0$ -- $2^3S_1$ transition, in cm$^{-1}$ unless specified:
}\\
12 &   958x48(1)     &    95x847   &    95x848     &    95x848    &   958x51(7)      &      \cite{klein:85}   \\
14 &  1138x09(2)     &   113x809   &   113x809     &   113x809    &  1138x07(4)      &      \cite{howie:94}   \\
   &                 &             &               &              &  1138x15(4)      &      \cite{deserio:81} \\
15 &  1229x56(2)     &   122x955   &               &   122x955    &  1229x53(9)      &      \cite{howie:94}   \\
16 &  1322x20(2)     &   132x219   &   132x219     &   132x219    &  1322x14(7)      &      \cite{howie:96}   \\
   &                 &             &               &              &  1321x98(10)     &      \cite{deserio:81} \\
18 &  1511x58(3)     &   151x155   &   151x156     &   151x155    &  1511x64(4)      &      \cite{kukla:95}   \\
   &                 &             &               &              &  1512x04(9)      &      \cite{beyer:86}   \\
26 &  2334x84(10)    &   233x469   &   233x471     &              &  2325x58(550)    &      \cite{buchet:81}  \\
36 &  3568x92(39)    &   356x822   &   356x828     &   356x823    &  3574x00(260)    &      \cite{martin:90}  \\
92 &  251.x93(26)\ {\rm eV} &   252x.79  &   252x.77   &          &  260.x0(7.9)     &      \cite{munger:86}  \\
\hline
\multicolumn{5}{l}{
$2^3P_2$ -- $2^3S_1$ transition, in cm$^{-1}$:
}\\
12 &  1002x53(1)   &     100x252   &   100x253     &   100x252    &    1002x63(6)      &    \cite{klein:85}      \\
14 &  1227x44(1)   &     122x743   &   122x743     &   122x743    &    1227x43(3)      &    \cite{howie:94}      \\
   &               &               &               &              &    1227x46(3)      &    \cite{deserio:81}    \\
15 &  1351x54(1)   &     135x151   &               &   135x151    &    1351x50(5)      &    \cite{howie:94}      \\
16 &  1484x99(1)   &     148x496   &   148x497     &   148x496    &    1484x98(4)      &    \cite{howie:96}      \\
   &               &               &               &              &    1484x93(5)      &    \cite{deserio:81}    \\
18 &  1785x82(1)   &     178x576   &   178x578     &   178x576    &    1785x89(5)      &    \cite{kukla:95}      \\
   &               &               &               &              &    1785x91(31)     &    \cite{beyer:86}      \\
20 &  2141x79(2)   &     214x170   &   214x174     &   214x170    &    2142x25(45)     &    \cite{hinterlong:86} \\
22 &  2566x96(2)   &     256x683   &   256x688     &   256x683    &    2567x46(46)     &    \cite{galvez:86}     \\
26 &  3687x67(6)   &     368x742   &   368x752     &   368x742    &    3689x76(125)    &    \cite{buchet:81}     \\
28 &  4419x42(8)   &     441x908   &   441x920     &   441x907    &    4419x50(80)     &    \cite{zacarias:88}   \\
36 &  9001x16(33)  &     900x009   &   900x044     &   900x008    &    9000x10(240)    &    \cite{martin:90}     \\
\hline
\multicolumn{5}{l}{
$2^3P_1$ -- $2^3S_1$ transition, in cm$^{-1}$:
}\\
12 &    966x82(1)    &    966x80   &    966x81     &              &   966x83(6)    &   \cite{klein:85} \\
13 &   1060x26(1)    &   1060x25   &               &              &  1060x23(7)    &   \cite{klein:85} \\
\hline
\multicolumn{5}{l}{
$2^3P_1$ -- $2^1S_0$ transition, in cm$^{-1}$:
}\\
14 &    722x9(1)     &    723x1    &               &              &   723x0.5(2)    &  \cite{redshaw:02}\\
\hline
\multicolumn{5}{l}{
$2^3P_0$ -- $2^3P_1$ transition, in eV unless specified:
}\\
12 & 83x4(1)\ {\rm cm}^{-1} &8x33    &    8x33        &             &  833.x133(15)      &     \cite{myers:01} \\
28 &    2.3x24(1)      &   2.x323   &   2.x325       &             &    2.x33(15)       & \cite{dunford:91}   \\
47 &    0.8x03(8)      &   0.x801   &   0.x789       &             &    0.x79(4)        & \cite{birkett:93}   \\
64 &   18.5x86(31)     &  18.x571   &  18.x548       &             &   18.x57(19)       & \cite{indelicato:92}\\
\hline
\multicolumn{5}{l}{
$1^1S_0$ -- $2^1P_1$ transition, in eV:
}\\
16  &   246x0.629      &    246x0.628    & & &   2460.x649(9)  &   \cite{schleinkofer:82}  \\
18  &   313x9.582      &    313x9.580    & & &   3139.x553(38) &   \cite{deslattes:84}     \\
19  &   351x0.461      &    351x0.459    & & &   3510.x58(12)  &   \cite{beiersdorfer:89}  \\
21  &   431x5.412      &    431x5.409    & & &   4315.x54(15)  &   \cite{beiersdorfer:89}  \\
22  &   474x9.644(1)   &    474x9.639    & & &   4749.x74(17)  &   \cite{beiersdorfer:89}  \\
23  &   520x5.165(1)   &    520x5.154    & & &   5205.x27(21)  &   \cite{beiersdorfer:89}  \\
    &                  &                 & & &   5205.x10(14)  &   \cite{chantler:00}      \\
24  &   568x2.068(1)   &    568x2.061    & & &   5682.x32(40)  &   \cite{beiersdorfer:89}  \\
26  &   670x0.434(1)   &    670x0.423    & & &   6700.x73(20)  &   \cite{beiersdorfer:89}  \\
    &                  &                 & & &   6700.x90(25)  &   \cite{briand:84}        \\
32  &  1028x0.217(4)   &   1028x0.185    & & &   10280x.70(22) &   \cite{maclaren:92}      \\
36  &  1311x4.470(7)   &   1311x4.411    & & &   13115x.31(30) &   \cite{indelicato:86}    \\
    &                  &                 & & &   13114x.68(36) &   \cite{widmann:96}       \\
54  &  3063x0.049(61)  &   3062x9.667    & & &   30629x.1(3.5) &   \cite{briand:89}        \\
92  & 10061x0.44(56)   &  10061x3.924    & & &   10062x6(35)   &   \cite{briand:90}        \\
\hline
\multicolumn{5}{l}{
$1^1S_0$ -- $2^3P_1$, in eV:
}\\
18  &   312x3.534       &   312x3.532    &  & &  3123.x522(36) &   \cite{deslattes:84}     \\
23  &   518x0.326(1)    &   518x0.327    &  & &  5180.x22(17)  &   \cite{chantler:00}      \\
26  &   666x7.578(1)    &   666x7.564    &  & &  6667.x50(25)  &   \cite{briand:84}        \\
32  &  1022x0.799(4)    &  1022x0.759    &  & &  10221x.80(35) &   \cite{maclaren:92}      \\
36  &  1302x6.116(7)    &  1302x6.044    &  & &  13026x.8(3)   &   \cite{indelicato:86}    \\
54  &  3020x6.263(61)   &  3020x5.852    &  & &  30209x.6(3.5) &   \cite{briand:89}        \\
92  &  9616x9.19(56)    &  9617x2.427    &  & &  96171x(52)    &   \cite{briand:90}        \\
\hline
\multicolumn{5}{l}{
$1^1S_0$ -- $2^3P_2$, in eV:
}\\
23  &  5188x.738(1)     &  5188x.730     &  & & 5189x.12(21)  &  \cite{chantler:00} \\
\hline
\multicolumn{5}{l}{
$1^1S_0$ -- $2^3S_1$, in eV:
}\\
23  &   5153x.896(1)    &   5153x.889    &  & & 5153x.82(14) &  \cite{chantler:00} \\
\colrule\hline
\end{longtable*}
\end{ruledtabular}
\endgroup

%%%%%%%%%%%%%%%%%%%%%%%%%%%%%%%%%%%

%%%%%%%%%%%%%%%%%%%%%%%%%%%%%%%%%%%
\begingroup
\squeezetable
\begin{table}
\caption{The $2\,^3P_0-2\,^1S_0$ transition energy, in eV.
 \label{tab:pncen}}
\begin{ruledtabular}
\begin{tabular}{l....|....}
& \multicolumn{1}{c}{$Z=63$} & \multicolumn{1}{c}{$Z=64$} &  \multicolumn{1}{c}{$Z=65$}
            &  \multicolumn{1}{c}{$Z=66$}    &  \multicolumn{1}{c}{ $Z=89$}   &  \multicolumn{1}{c}{$Z=90$}
                   &   \multicolumn{1}{c}{$Z=91$}  &    \multicolumn{1}{c}{$Z=92$}    \\
\colrule
This work                       & -0.22x6(73) &   0.00x6(79)&  0.29x6(84) & 0.49x3(91)  &    1.6x1(46) &  0.0x1(25) & -0.3x1(54) &   -2.7x1(27)  \\
Andreev \etal \cite{andreev:03} & -0.59x1     &  -0.38x9    & -0.15x3     & 0.01x6      &              &            & -1.9x71    &   -4.5x11     \\
Plante  \etal \cite{plante:94}  &             &  -0.17x0    &             & 0.34x1      &              & -0.0x95    &            &   -2.6x39     \\
Drake   \cite{drake:CJP:88}     & -0.16x8     &   0.06x7    &  0.32x8     & 0.61x4      &    1.7x31    &  0.7x18    & -0.2x09    &   -1.8x16     \\
Maul \etal \cite{maul:96}       &             &             &             &             &              &            &            &    0.3x0      \\
\end{tabular}
\end{ruledtabular}
\end{table}
\endgroup

%%%%%%%%%%%%%%%%%%%%%%%%%%%%%%%%%%%%%%%%%%%%%%%%%%%%%%%%%%%%%%%%%%%%%%%%
%%%%%
%%%%%
%%%%%%%%%%%%%%%%%%%%%%%%%%%%%%%%%%%%%%%%%%%%%%%%%%%%%%%%%%%%%%%%%%%%%%%
%\newpage
\begin{figure}
\centering
\includegraphics[clip=true,width=0.2\textwidth]{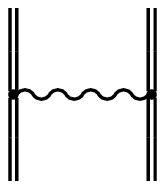}
\caption{The diagram of the one-photon exchange. \label{fig:onephot}}
\end{figure}

%%%%%%%%%%%%%%%%%%%%%%%%%%%%%%%%%%%%%%%%%%%%%%%%%%%%%%%%%%%%%%%%%%%%%%%%
%%%%%
%%%%%
%%%%%%%%%%%%%%%%%%%%%%%%%%%%%%%%%%%%%%%%%%%%%%%%%%%%%%%%%%%%%%%%%%%%%%%
\newpage
\begin{figure}
\centering
\includegraphics[clip=true,width=0.5\textwidth]{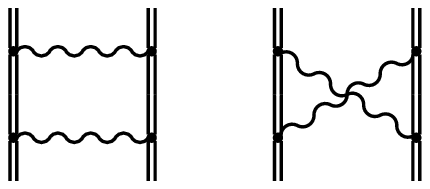}
\caption{ The diagrams of the two-photon exchange. \label{fig:twophot}}
\end{figure}

%%%%%%%%%%%%%%%%%%%%%%%%%%%%%%%%%%%%%%%%%%%%%%%%%%%%%%%%%%%%%%%%%%%%%%%%
%%%%%
%%%%%
%%%%%%%%%%%%%%%%%%%%%%%%%%%%%%%%%%%%%%%%%%%%%%%%%%%%%%%%%%%%%%%%%%%%%%%
\newpage
\begin{figure}
\centering
\includegraphics[clip=true,width=0.9\textwidth]{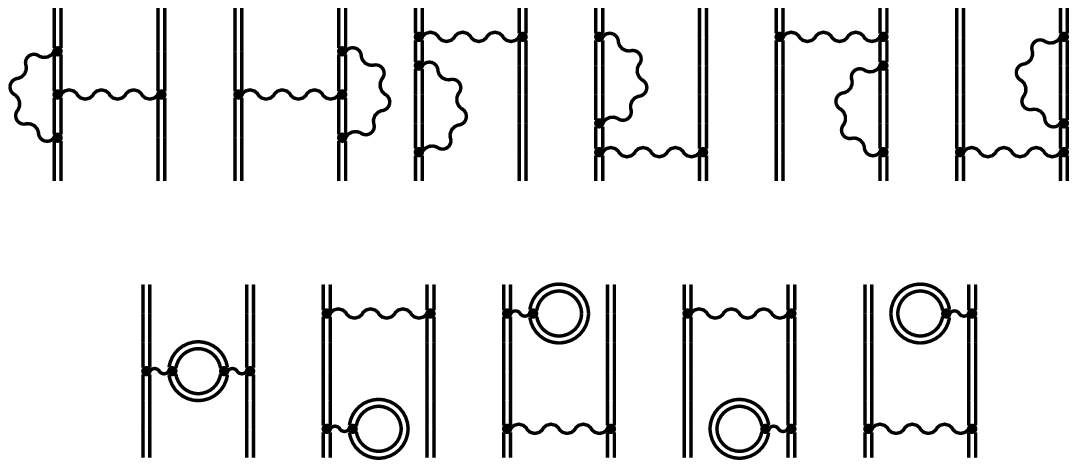}
\caption{ Self-energy screening and vacuum-polarization screening diagrams.
 \label{fig:sevpscr}}
\end{figure}

%%%%%%%%%%%%%%%%%%%%%%%%%%%%%%%%%%%%%%%%%%%%%%%%%%%%%%%%%%%%%%%%%%%%%%%%
%%%%%
%%%%%
%%%%%%%%%%%%%%%%%%%%%%%%%%%%%%%%%%%%%%%%%%%%%%%%%%%%%%%%%%%%%%%%%%%%%%%
\newpage
\begin{figure*}
\centering
\includegraphics[clip=true,width=\textwidth]{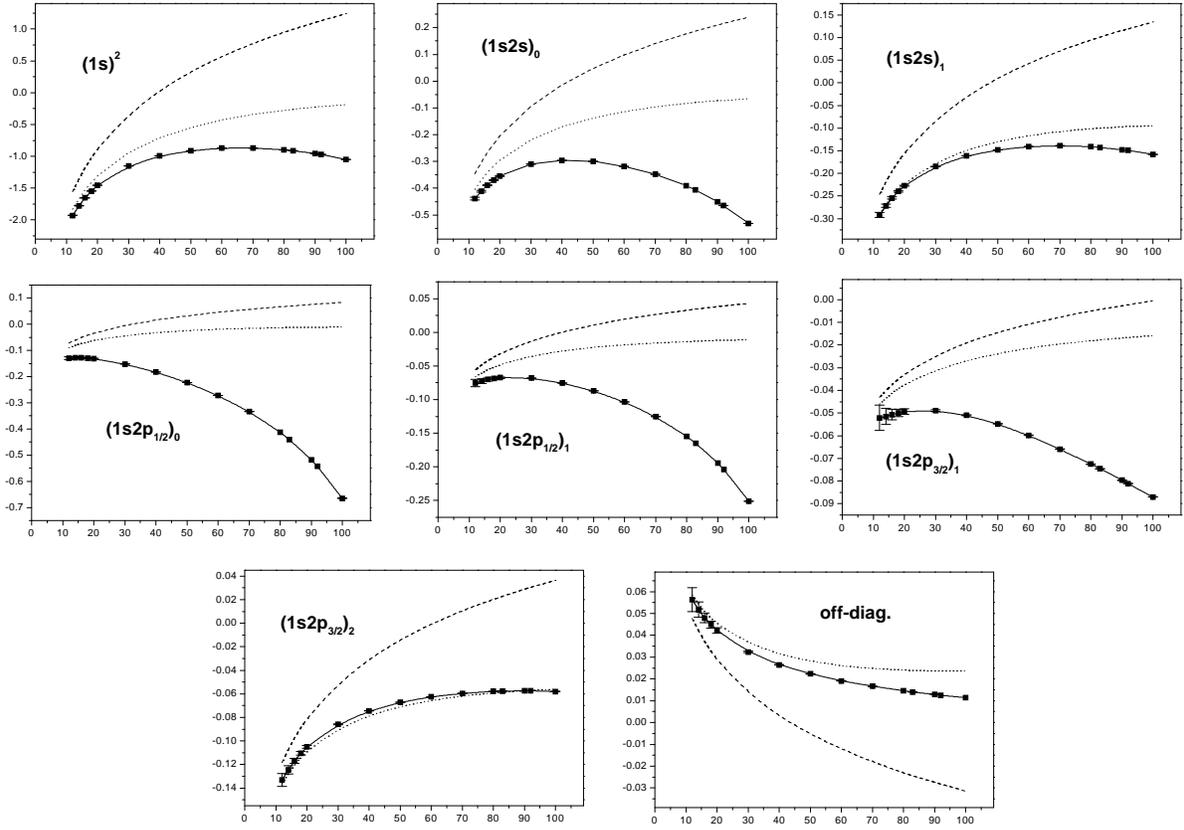}
\caption{Comparison of our all-order numerical results for the
second-order two-electron
QED correction (square dots, solid line) with values for this
correction within the $\aZ$ expansion (the contribution of order $\alpha^2
(\aZ)^3$, dashed line) and with the related QED contribution by Drake
\cite{drake:CJP:88} (dotted line), in units of $\alpha^2(\aZ)^3$.
\label{fig:comp} }
\end{figure*}

%%%%%%%%%%%%%%%%%%%%%%%%%%%%%%%%%%%%%%%%%%%%%%%%%%%%%%%%%%%%%%%%%%%%%%%%
%
%
%
%%%%%%%%%%%%%%%%%%%%%%%%%%%%%%%%%%%%%%%%%%%%%%%%%%%%%%%%%%%%%%%%%%%%%%%%
\newpage
\begin{figure*}
\centering
\includegraphics[clip=true,width=1.1\textwidth]{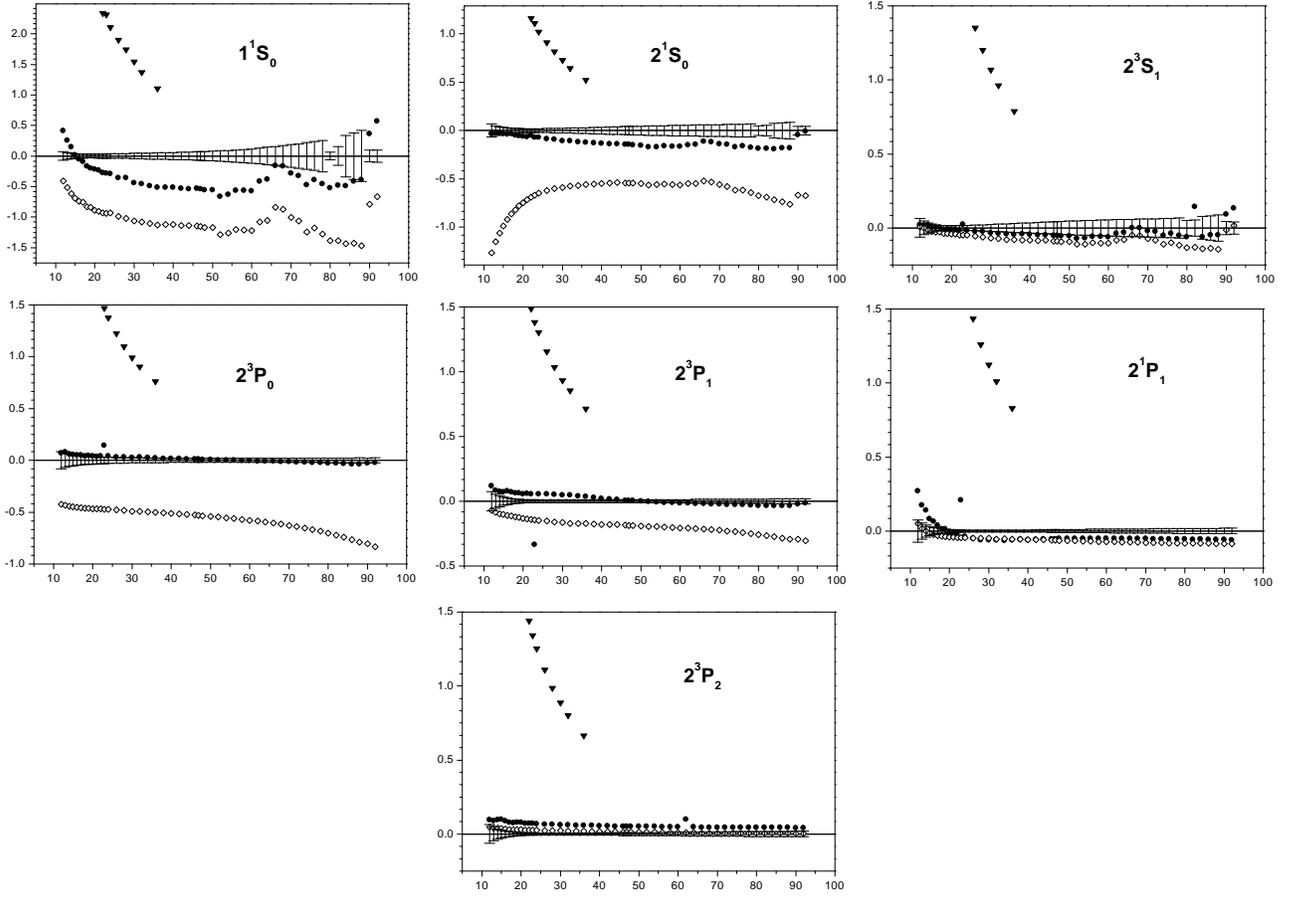}
\caption{Comparison of different evaluations for the total ionization energy
of $n=1$ and $n=2$ states of He-like ions. Plotted is the difference between
the results obtained by us and by other authors, normalized by the factor
$\alpha^2\,(\aZ)^4$. Error bars refer to the estimation of the
uncertainty of the present evaluation, open diamonds denote the results by
Drake \cite{drake:CJP:88}, filled circles stand for those by Plante {\it et
al.} \cite{plante:94}, and filled triangles indicate the values by Cheng and
Chen \cite{cheng:00}. \label{fig:totcomp} }
\end{figure*}

%\bibliographystyle{/home/2/iftguest/yerokhin/papers/phaip30}
%\bibliography{/home/2/iftguest/yerokhin/papers/hfst}

\end{document}